\documentstyle[12pt]{article}
\input{amssym.def}
\input{amssym}

\def\theequation{\thesection.\arabic{equation}}

\def\be{\begin{equation}}
\def\ee{\end{equation}}
\def\ba{\begin{eqnarray}}
\def\ea{\end{eqnarray}}
\def\lb{\label}
\def\nin{\noindent}

\begin{document}
\begin{center}
\vskip .5cm

{\LARGE{\bf Raise and Peel Models  of fluctuating interfaces
and  combinatorics of Pascal's\rule{0pt}{7mm} hexagon\rule{0pt}{7mm}}}

\vskip .6cm

{\large {\bf P. Pyatov}\footnote{E-mail: pyatov@thsun1.jinr.ru}
}

\vskip 0.2 cm

Bogoliubov Laboratory of Theoretical Physics, JINR \\
141980 Dubna, Moscow Region, Russia

\end{center}

\vspace{.5 cm} \nin {\sc Abstract.} {\sf

The raise and peel model of a one-dimensional fluctuating interface
(model A) is extended by considering one source (model B) or two
sources (model C) at the boundaries. The Hamiltonians describing
the three processes have, in the thermodynamic limit, spectra given
by conformal field theory.
The probability of the different configurations in the stationary
states of the three models are not only related but have interesting
combinatorial properties. We show that by extending Pascal's triangle
(which gives solutions to linear relations in terms of integer numbers),
to an hexagon, one obtains integer solutions of bilinear relations. These
solutions give not only the weights of the various configurations in the
three models but also give an insight to the connections between the probability
distributions in the stationary states of the three models. Interestingly
enough, Pascal's hexagon also gives solutions to a Hirota's difference
equation.
}

\section{Introduction.}
\setcounter{equation}0

Recently much interest have been devoted to surprising
appearence of the ASM (alternating sign matrices) combinatorics
in the properties of the ground state wave function of XXZ spin chain at a
particular value $\Delta=-1/2$  of its anisotropy \cite{RS1}. In
subsequent investigations \cite{BGN1}--\cite{GNPR2} a number of
models having ground states with interesting
combinatorial properties was found,
including the dense O(1) loop model, the rotor model and the
"raise and peel" model of fluctuating interface. Two important
facts are common for these models. First, all they admit a purely
algebraic description in terms of an appropriate version of a
(quotient of) Temperley--Lieb (TL) algebra (see \cite{PRGN}).
Secondly, the TL algebras always come in a very specific semigroup
regime. This latter fact was used in \cite{PRG,PRGN} to
interpret the loop model as a stochastic process and thus, to give a
physical interpretation to the components of the ground
state wave function  as (unnormalized)
probabilities of various configurations.
\medskip

In the present paper we extend the raise and peel model\footnote{
For detailed discussion of physical properties of
the raise and peel model the reader is referred to papers \cite{GNPR2,LR}.}
(hereafter called model A) obtained for the TL algebra
to two other cases.
In order to do so, we consider the boundary extension of the TL
algebra which is called the blob algebra \cite{MS,MW1,G} (see Section 2).
In this way, in the stochastic model one can introduce a source at
one boundary (model B) or sources at the two boundaries (model C).
The two new models are described in detail in Section 3. As shown
in Ref.\cite{GNiPR}, in the continuum limit, the spectra of the two
Hamiltonians giving the time evolution of the stochastic processes
are given by characters of N=2 superconformal field theory. Here
we are going to consider the combinatorial properties of the stationary
states only.

In Section 4, based on numerical studies of small chains we make a
series of conjectures for the weights of various configurations
observed in the stationary states of the three models. In
Appendix A we define Pascal's hexagon. We think that the
content of this Appendix is interesting on its own. Using Pascal's hexagon one can get in a simple way the numerical results
obtained for the stationary states for finite systems (see
Section 4).

The Pascal's hexagon is connected in a profound (although not yet
understood way) to alternating sign matrices with various
symmetries and as discussed in Appendix A with the solutions of a
discrete Hirota's equation.
\medskip

After this work was almost completed, we learned from Jan de
Gier that part of our results were obtained independently by
Mitra et al \cite{MNGB}. We have also verified, as discussed in the
Appendix, that Pascal's hexagon can be used to obtain properties
of the stationary states of stochastic processes related to
the periodic TL algebra discussed in \cite{MNGB}.

\section{Boundary extended Temperley-Lieb algebra}
\setcounter{equation}0

\subsection{Definition.}

We start with the type $A$ Temperley-Lieb (TL) algebra \cite{TL}
which as it is well known (see \cite{PS}) stands
behind the $U_q(sl_2)$ symmetric
XXZ spin chain. For the chain of $L$ particles
the corresponding TL algebra
is generated by the unity and
a set of $(L-1)$  elements $e_i$, $i=1,\dots (L-1)$,
subject to relations
\ba
\lb{TL1}
e_i e_{i\pm 1} e_i = e_i\ ,&& e_i^2 = (q+q^{-1})\, e_i\ ,
\\[2mm]
\lb{TL2} e_i e_j = e_j e_i\ , &&\qquad \quad \forall\ i,j:
|i-j|>1\ . \ea Here  parameter $q\in {\Bbb C}\backslash \{0\}$ of
the algebra is related to the anisotropy parameter
$\Delta=-(q+q^{-1})/2$ of the spin chain.

A boundary extension of this algebra is achieved by adding two
more generators $f_0$ and $f_L$ together with the relations \ba
\lb{TL3} e_1 f_0 e_1 = e_1\ ,&& e_j f_0 = f_0 e_j\ , \;\quad
\forall\; j > 1 ,
\\[1mm]
\lb{TL4}
e_{L-1} f_L e_{L-1} = e_{L-1}\ ,&& e_j f_L = f_L e_j\ , \;\quad \forall\; j < L-1 .
\\[2mm]
\lb{TL5}
f_0^2 =  a\,  f_0\ ,\qquad f_L^2 =  {\bar a}\, f_L\ , &&
f_0 f_L = f_L f_0\ ,
\ea
where $a, {\bar a} \in {\Bbb C}$.
The algebra with one boundary element (either $f_0$, or $f_L$)
called the {\em blob algebra} was analyzed in
\cite{MS,MW1}.
The  extension of the TL algebra with the two boundary generators
was introduced in \cite{G}.
Unlike TL and blob algebras it is infinite dimensional
and we are going to further extract its finite
dimensional quotient. To this end we consider a pair of
unnormalized projectors $X_L$ and $Y_L$. They are defined differently
depending on a parity of $L$
\be
\begin{array}{lll}
\mbox{for $L$ even:}&
 X_L \ :=\ {\displaystyle \prod_{k=0}^{L/2-1}} e_{2k+1}\ , \;\;\;&
 Y_L \ :=\ f_0 {\displaystyle \prod_{k=1}^{L/2-1}} e_{2k}\ f_L\ ;\;\;\;
\\
\\
\mbox{for $L$ odd:}&
X_L \ :=\ f_0 {\displaystyle \prod_{k=1}^{(L-1)/2}} e_{2k}\   , \;\;\;&
Y_L \ :=\ {\displaystyle \prod_{k=0}^{(L-3)/2}} e_{2k+1}\ f_L\ .\;\;\;
\end{array}
\ee
In terms of these projectors  reduction conditions  read
\be
\lb{red}
X_L Y_L X_L = b\ X_L\ , \quad Y_L X_L Y_L = b\ Y_L\ .
\ee
The resulting quotient algebra is that one we shall further call
the {\em boundary extended TL algebra}. It is finite
dimensional\footnote{ Dimensions of the boundary extended TL
algebras are calculated in an  Appendix to \cite{GP}.} and it
depends on four parameters $q$, $a$, $\bar a$ and $b\in {\Bbb C}$.

\subsection{Graphical presentation.}

There are at least two ways in which the boundary extended TL
algebra can be visualized.
First one is a straightforward generalization of
the diagrammatic realization of the blob algebra presented in
\cite{MS}.
For this one uses familiar ''lines and loops'' diagrams
for the TL generators $e_i$
and realizes the boundary generators $f_0$ and $f_L$ as two different
blobs lying, respectively, on the leftmost and the rightmost lines
of the diagram (see \cite{MS}). Equivalently, one can draw boundary
generator $f_0$ ($f_L$) as a half-loop connecting the leftmost (rightmost)
line to the boundary (see \cite{G,MNGB}).

The second way which we are using throughout this paper is the one
suitable for modelling of growing interfaces (see \cite{GNPR2}).
One draws the TL generator $e_i$ as a tile whose diagonal is lying
on a vertical line with coordinate $i$ and whose left and right
vertices are placed, respectively, on vertical lines with
coordinates $i-1$ and $i+1$. The boundary generators $f_0$ and
$f_L$ are drawn as half-tiles with their longest sides lying on
vertical lines with coordinates $0$ and $L$, respectively (see
Figure below). \be \lb{graph}
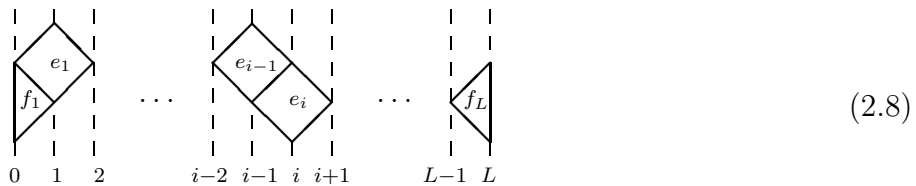
\begin{array}{c}
\begin{picture}(50,70)(0,-10)
\linethickness{0.25pt}
\multiput(-80,60)(0,-10){6}{\line(0,-1){5}}
\multiput(-65,60)(0,-10){1}{\line(0,-1){5}}
\multiput(-65,20)(0,-10){2}{\line(0,-1){5}}
\multiput(-50,60)(0,-10){6}{\line(0,-1){5}}
{\thicklines
\put(-80,40){\line(1,-1){15}}
\put(-80,40){\line(1,1){15}}
\put(-80,10){\line(1,1){30}}
\put(-65,55){\line(1,-1){15}}
\put(-80,10){\line(0,1){30}}
\put(-78,24){$\scriptstyle f_1$}
\put(-66.5,38){$\scriptstyle e_1$}
}
\multiput(-5,60)(0,-10){6}{\line(0,-1){5}}
\multiput(10,60)(0,-10){1}{\line(0,-1){5}}
\multiput(10,20)(0,-10){2}{\line(0,-1){5}}
\multiput(25,60)(0,-10){2}{\line(0,-1){5}}
\multiput(25,10)(0,-10){1}{\line(0,-1){5}}
\multiput(40,60)(0,-10){6}{\line(0,-1){5}}
{\thicklines
\put(10,25){\line(1,1){15}}
\put(-5,40){\line(1,-1){30}}
\put(10,55){\line(1,-1){30}}
\put(-5,40){\line(1,1){15}}
\put(25,10){\line(1,1){15}}
\put(24,24){$\scriptstyle e_i$}
\put(3.5,38){$\scriptstyle e_{i-1}$}
}
\multiput(85,60)(0,-10){6}{\line(0,-1){5}}
\multiput(100,60)(0,-10){6}{\line(0,-1){5}}
{\thicklines
\put(85,25){\line(1,1){15}}
\put(85,25){\line(1,-1){15}}
\put(100,10){\line(0,1){30}}
\put(90,24){$\scriptstyle f_L$}
}
\put(-33,25){\dots}
\put(57,25){\dots}
\put(-82,-5){$\scriptstyle 0$~~~$\scriptstyle 1$~~~$\scriptstyle 2$%
~~~~~~~~$\scriptstyle \, i-2$~%
$\scriptstyle \, i-1$~$\scriptstyle \, i$~$\scriptstyle \, i+1$%
~~~~~~~$\scriptstyle L-1$~$\scriptstyle \, L$}
\end{picture}
\end{array}
\ee The (half-)tiles can freely move along vertical axes unless
they meet their neighbors. Assuming attraction forces acting among
the (half-)tiles one represents word in the algebra as a
collection of dense polygons built from the (half-)tiles and
satisfying following conditions. All polygons are placed  between
vertical lines with coordinates $0$ and $L$ and no vertical line
lying between these two boundary verticals crosses the borders of
(one or several) polygons in more then two points.

\subsection{The ideal ${\cal I}_L$.}

Of our main interest is the left ideal in the boundary extended TL algebra
generated by $X_L$\footnote{
Note that in case $b\neq 0$ the element $Y_L$ generates an isomorphic ideal.
}. We denote this ideal as ${\cal I}_L$.

Consider graphical realization of a typical word in the ideal.
As we have different definitions of $X_L$ depending on a parity of $L$,
separate  pictures for the cases of $L$ even and $L$ odd are
given below.
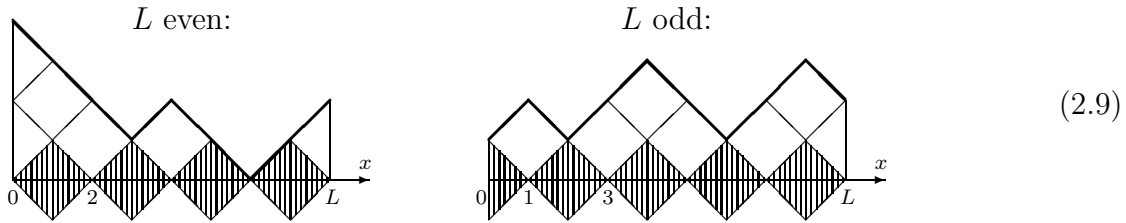
\be
\lb{RSOS}
\begin{array}{c}
\begin{picture}(150,100)(-70,-20)
\put(-135,57){$L$ even:}
\put(50,57){$L$ odd:}
{\linethickness{0.4pt}
\put(-180,0){\vector(1,0){135}}
\put(0,0){\vector(1,0){150}}
}
{\linethickness{0.4pt}
\multiput(-163,-13)(30,0){4}{\line(0,1){26}}
\multiput(-161,-11)(30,0){4}{\line(0,1){22}}
\multiput(-159,-9)(30,0){4}{\line(0,1){18}}
\multiput(-157,-7)(30,0){4}{\line(0,1){14}}
\multiput(-155,-5)(30,0){4}{\line(0,1){10}}
\multiput(-153,-3)(30,0){4}{\line(0,1){6}}
\multiput(-151,-1)(30,0){4}{\line(0,1){2}}
\multiput(-179,-1)(30,0){4}{\line(0,1){2}}
\multiput(-177,-3)(30,0){4}{\line(0,1){6}}
\multiput(-175,-5)(30,0){4}{\line(0,1){10}}
\multiput(-173,-7)(30,0){4}{\line(0,1){14}}
\multiput(-171,-9)(30,0){4}{\line(0,1){18}}
\multiput(-169,-11)(30,0){4}{\line(0,1){22}}
\multiput(-167,-13)(30,0){4}{\line(0,1){26}}
\multiput(-165,-15)(30,0){4}{\line(0,1){30}}
}
\multiput(-180,0)(30,0){4}{\line(1,1){15}}
\multiput(-165,-15)(30,0){4}{\line(1,1){15}}
\multiput(-180,0)(15,15){2}{\line(1,-1){15}}
\multiput(-150,0)(15,15){2}{\line(1,-1){15}}
\multiput(-120,0)(15,15){2}{\line(1,-1){15}}
\multiput(-90,0)(15,15){2}{\line(1,-1){15}}
\put(-182,-9){$\scriptstyle 0$}
\put(-152,-9){$\scriptstyle 2$}
\put(-62,-9){$\scriptstyle L$}
\put(-49,4){$\scriptstyle x$}
\put(-165,15){\line(1,1){15}}
\put(-180,30){\line(1,1){15}}
\put(-180,60){\thicklines\line(1,-1){45}}
\put(-180,60.5){\thicklines\line(1,-1){45}}
\put(-180,30){\line(1,-1){15}}
\put(-180,0){\line(0,1){60}}
\put(-135,15){\thicklines\line(1,1){15}}
\put(-135,15.5){\thicklines\line(1,1){15}}
\put(-120,30){\thicklines\line(1,-1){30}}
\put(-120,30.5){\thicklines\line(1,-1){30}}
\put(-90,0){\thicklines\line(1,1){30}}
\put(-90,0.5){\thicklines\line(1,1){30}}
\put(-60,0){\line(0,1){30}}
\multiput(0,-15)(30,0){5}{\line(1,1){15}}
\multiput(15,0)(15,15){2}{\line(1,-1){15}}
\multiput(45,0)(15,15){2}{\line(1,-1){15}}
\multiput(75,0)(15,15){2}{\line(1,-1){15}}
\multiput(105,0)(15,15){2}{\line(1,-1){15}}
\put(-5,-9){$\scriptstyle 0$}
\put(13,-9){$\scriptstyle 1$}
\put(43,-9){$\scriptstyle 3$}
\put(133,-9){$\scriptstyle L$}
\put(146,4){$\scriptstyle x$}
{\linethickness{0.4pt}
\multiput(2,-13)(30,0){5}{\line(0,1){26}}
\multiput(4,-11)(30,0){5}{\line(0,1){22}}
\multiput(6,-9)(30,0){5}{\line(0,1){18}}
\multiput(8,-7)(30,0){5}{\line(0,1){14}}
\multiput(10,-5)(30,0){5}{\line(0,1){10}}
\multiput(12,-3)(30,0){5}{\line(0,1){6}}
\multiput(14,-1)(30,0){5}{\line(0,1){2}}
\multiput(16,-1)(30,0){4}{\line(0,1){2}}
\multiput(18,-3)(30,0){4}{\line(0,1){6}}
\multiput(20,-5)(30,0){4}{\line(0,1){10}}
\multiput(22,-7)(30,0){4}{\line(0,1){14}}
\multiput(24,-9)(30,0){4}{\line(0,1){18}}
\multiput(26,-11)(30,0){4}{\line(0,1){22}}
\multiput(28,-13)(30,0){4}{\line(0,1){26}}
\multiput(30,-15)(30,0){4}{\line(0,1){30}}
}
\put(0,-15){\line(0,1){30}}
\put(0,15){\line(1,-1){15}}
\multiput(15,0)(30,0){4}{\line(1,1){15}}
\put(120,15){\line(1,1){15}}
\put(90,15){\thicklines\line(1,1){30}}
\put(90,15.5){\thicklines\line(1,1){30}}
\put(105,30){\line(1,-1){15}}
\put(120,45){\thicklines\line(1,-1){15}}
\put(120,45.5){\thicklines\line(1,-1){15}}
\put(135,0){\line(0,1){30}}
\put(60,15){\line(1,1){15}}
\put(30,15){\thicklines\line(1,1){30}}
\put(30,15.5){\thicklines\line(1,1){30}}
\put(45,30){\line(1,-1){15}}
\put(60,45){\thicklines\line(1,-1){30}}
\put(60,45.5){\thicklines\line(1,-1){30}}
\put(0,15){\thicklines\line(1,1){15}}
\put(0,15.5){\thicklines\line(1,1){15}}
\put(15,30){\thicklines\line(1,-1){15}}
\put(15,30.5){\thicklines\line(1,-1){15}}
\end{picture}
\end{array}
\ee
Here we adopt a convention that multiplication from the left
by elements $e_i$, $f_0$, or $f_L$ amounts graphically to dropping
their respective (half-)tiles up-down. Components of $X_L$ are
shown hatched on the pictures.

As it is obvious from pictures (\ref{RSOS}) each word $w$ in the
ideal ${\cal I}_L$ is uniquely defined by a shape of the upper
border $h(w|x)$, $0\leq x\leq L$, of its corresponding polygon
(drawn in bold lines on the pictures). In turn, the  border line
is suitably encoded by its values at the integer points
$h(w|i):=h_i(w)$. $i=0,1,\dots ,L$. Assuming the height of the
tile ($=$ the length of its diagonal) equals 2 and taking the
middle line of the bottom row of tiles as a reference axe one gets
following prescriptions for a set of $\{h_i\}_{i=0,1,\dots ,L}$
\ba
\nonumber
&a)&
h_{i+1}-h_i = \pm 1 \quad \mbox{\em and}\quad  h_i\geq 0\, ,
\;\forall\; i\, ; \hspace{8cm}
\\[2mm]
\lb{hi}
&b)&
h_L \mbox{\em ~is an even integer;}
\\[2mm]
\nonumber
&c)&
\mbox{\em there exists $i$ such that~ }  h_i\in\{0,1\}\, .
\ea
Here prescription b) results from our choice of $X_L$ as
an ideal generating element.
With the choice of $Y_L$ one would constrain $h_0$ to be even.
Prescription c) arises from the reduction conditions (\ref{red}).

In Ref.\cite{BE} a set of data $\{h_i\}_{i=0,1,\dots ,L}$ satisfying conditions
(\ref{hi}) is named an {\em Anchored Cross path},
$L$ is called a {\em length} of the path.
Anchored Cross paths of
length $L$ label effectively words in the ideal ${\cal I}_L$.
There are $2^L$ different Anchored Cross paths of  length $L$
(for the proof c.f. Appendix of Ref.\cite{GP})
and thus, $\dim {\cal I}_L = 2^L$.
It is remarkable  that the dimension of
${\cal I}_L$ coincides with the number of states of the
chain of $L$ spin=1/2 particles.
This is not just a coincidence
and the ideal ${\cal I}_L$ can be used for representation
of an open XXZ chain of $L$ spin=1/2 particles (see \cite{GNiPR}).

In considerations below we will use
besides the set of Anchored Cross paths
a pair of its subsets
(or, equivalently, two subspaces in the ideal ${\cal I}_L$).
Their definitions are given below.

{\em Ballot paths} are the paths (\ref{hi}) with fixed endpoint
$h_L=0$. Their total number is ${L \choose [L/2]}$, where $[x]$ is
an integer part of $x$. Examples of Ballot paths are shown on
pictures a), b) and c) on
Fig.\ref{absorb-desorb}
on page \pageref{absorb-desorb}. Paths shown on pictures d) and e)  are
not the Ballot paths.

{\em Dyck paths} are usually defined for $L$
even and they are fixed at both ends as $h_0=h_L=0$. For $L=2p$ one
has $C_p := {1\over p+1}{2p \choose p}$ Dyck paths which is the
$p$-th Catalan number.
For $L$ odd close relatives of Dyck
paths are those whose endpoints are fixed as $h_0=1$, $h_L=0$.
These paths are in one to one correspondence with the Dyck paths
of length $L+1$ and later on we will also refer them as Dyck
paths. Among the paths shown on Fig. \ref{Fig1a} on page
\pageref{Fig1a} cases a), b) and c) are the Dyck paths, while
cases d), e), f) are not.

\section{Raise and Peel Models with different boundary terms.}
\setcounter{equation}0

\subsection{The models definition.}

First, we describe the models algebraically and then,
we  discuss their physical interpretation.

\medskip
By definition, the ideal ${\cal I}_L$ forms  left
representation space of the boundary extended TL algebra.
Consider on this space a dynamical process
\ba
\lb{evol}
{d\over dt}\, | p_L(t){\cal i}& =& -\, H_L\,
|p_L(t){\cal i}\, ,
\\
\lb{H}
H_L &:=& \sum_{i=1}^{l-1}(1-e_i) + c(1-f_0) + {\bar c}(1-f_L)\, ,
\ea
defining an evolution of element $|p_L(t){\cal i}\in {\cal I}_L$.
Here $H_L$, the Hamiltonian of the process
contains two numeric parameters --- $c$ and $\bar c$, while the
process itself depends also on four parameters of the algebra ---
$q$, $a$, $\bar a$ and $b$
(see Eqs. (\ref{TL1}) -- (\ref{TL5}) and (\ref{red})).
We are interested in case where the boundary extended TL algebra
becomes semigroup
(that is,
all the nonvanishing structure constants of the algebra are units)
and so we fix algebra parameters as
\be
\lb{semigroup}
q=exp(i\pi/3)\; (\Rightarrow q+q^{-1}=1)\, , \quad a={\bar a}=b=1\, .
\ee
In this case the Hamiltonian (\ref{H}) becomes an intensity matrix
and the process (\ref{evol}) can be given a stochastic
interpretation (see, e.g., \cite{PRGN}).
Expanding element $|p_L(t){\cal i}$
into linear combination of words of the ideal
$$
|p_L(t){\cal i}=\sum_{w\in {\cal I}_L}\, p_L(w|t)\, w
$$
one treats coefficients $p_L(w|t)$ as
unnormalized probabilities to find the stochastic system in
configuration $w$ at time $t$.

In this paper we
consider  stochastic processes (\ref{evol}) corresponding
to three particular choises of
parameters $c$ and $\bar c$ of the Hamiltonian (\ref{H}).
We call them models A, B, and C,
\be
\lb{3cases}
\mbox{model A:~} c={\bar c}=0\, ; \qquad
\mbox{model B:~} c=1\, ,\;{\bar c}=0\, ; \qquad
\mbox{model C:~} c={\bar c}=1\, .
\ee
In cases A and B the Hamiltonian acts invariantly
on the subspaces of ${\cal I}_L$  spanned, respectively, by all Dyck and
Ballot paths. Therefore we shall treat models A/B on their
respective irreducible spaces of Dyck/Ballot paths.

\medskip
Now let us discuss physical interpretation of the models.
We consider three processes of growth of a film of tiles which
are deposited on a one-dimensional substrate of size $L$.

As a substrate in all cases we choose  profiles which are shown hatched on
pictures (\ref{RSOS}).
A rarefied gas above the substrate contains tiles and (possibly) half-tiles.
They are moving along  integer vertical lines as illustrated
on picture (\ref{graph}) and upon hitting the substrate they can be absorbed
and form interface configurations as shown on picture (\ref{RSOS}).
Depending on a
composition of the gas one distinguishes three cases
\begin{itemize}
\item[]model A:
the gas contains tiles moving along lines with coordinates
$i=1,2,\dots ,L-1$; possible interface configurations are given by
Dyck paths;
\item[]  model B:
the gas contains all the tiles and the half-tile moving along 0-th line;
possible interface configurations are the Ballot paths;
\item[] model C:
the gas contains all the tiles and the half-tiles on both left and
right boundaries; possible interface configurations are the
Anchored Cross paths.
\end{itemize}

To determine evolution rules in the models
we use the graphical presentation of the boundary extended
TL algebra.
We remind that in this presentation
the substrate of  size $L$ corresponds to
the unnormalized projector $X_L$;
interface configurations correspond to
words in  the ideal  ${\cal I}_L$;
the (half-)tile on $i$-th vertical line is an equivalent of the
algebra generator $e_i$ ($f_0$/$f_L$ for $i=0$/$L$);
hitting the interface by (half-)tiles amounts to
left multiplication by $e_i$ ($f_0$, $f_L$) in the ideal.
With these identifications
equation (\ref{evol}) defines following evolution rules.

During an infinitesimal time interval $dt$ a single event may
happen with equal probability rate at any integer point of the
interface. The following events are possible.
\begin{itemize}
\item[a)]
At a local minimum point $i$ (that is, if  $h_i < h_{i\pm 1}$)
the interface either absorbs \mbox{(half-)tile} ($h_i\mapsto h_{i+2}$)
with probability $dt$ or it reflects (half-)tile
($h_i$ stays unchanged) with
probability $1-dt$. For the model C there is
an exception from this rule described in item d).
\item[b)]
At a local maximum point $i$  (that is, if  $h_i > h_{i\pm 1}$) the
interface always reflects \mbox{(half-)tiles} and stays unchanged.
\item[c)]
At a bulk slope point $i$ (that is, $0<i<L$ and either
$h_{i-1}<h_i<h_{i+1}$, or $h_{i-1}>h_i>h_{i+1}$) dropping a tile
leads with probability $dt$ to a nonlocal desorption event called
avalanche. To describe the avalanche one determines integer $k$
such that  for all integers $j$ standing between $i$ and $k$
inequality $h_j>h_i$ holds and  either $h_k=h_i$, or $k$ runs out
the interval $[0,L]$, i.e., $k$ equals $L+1$, or $-1$. The
avalanche causes desorption of one tile at each point $j$ between
$i$ and $k$, that is $h_j\mapsto h_j -2$. The avalanche size (a
number of the desorbed (half-)tiles) $n_d=|i-k|-1$, $1\leq n_d\leq
L-1$, measures non-locality of the event.

With probability
$1-dt$ the tile is reflected and the interface stays unchanged.
\item[d)]
In the model C at a global minimum point $i$ such that $h_i=1$ and
$h_j>h_i\, ,\; \forall\; j\neq i\, ,$  dropping (half-)tile
with probability $dt$ causes  total avalanche of a size $n_d=L$
that is, $h_j\mapsto h_j-2\, ,\; \forall\; j\neq i$. Again, with
probability $1-dt$ the (half-)tile is reflected and the interface
stays unchanged.
\end{itemize}

Typical absorption and desorption events are illustrated
on Figure \ref{absorb-desorb}.

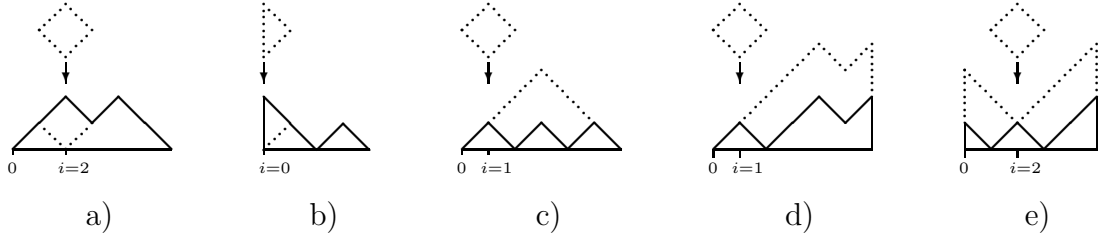
\begin{figure}
\hfil
$
\begin{array}{ccccc}
\begin{picture}(85,20)(0,0)
\put(0,0){\line(1,0){60}}
{\thicklines
\put(0,0){\line(1,1){20}}\put(20,20){\line(1,-1){10}}
\put(30,10){\line(1,1){10}}\put(40,20){\line(1,-1){20}}
}
\multiput(10,45)(2,2){5}{\circle*{.5}}\multiput(10,45)(2,-2){5}{\circle*{.5}}
\multiput(20,35)(2,2){5}{\circle*{.5}}\multiput(20,55)(2,-2){6}{\circle*{.5}}
\multiput(12,8)(2,-2){5}{\circle*{.5}}
\multiput(22,2)(2,2){5}{\circle*{.5}}
\put(0,0){\line(0,-1){2}}\put(-2,-9){$\scriptscriptstyle 0$}
\put(20,0){\line(0,-1){2}}\put(17,-9){$\scriptscriptstyle i=2$}
%
\put(20,33){\vector(0,-1){8}}
\end{picture}
&
\begin{picture}(65,20)(0,0)
\put(0,0){\line(1,0){40}}
\put(0,0){\line(0,1){20}}
{\thicklines
\put(0,20){\line(1,-1){20}}\put(20,0){\line(1,1){10}}
\put(30,10){\line(1,-1){10}}
}
\multiput(0,55)(2,-2){5}{\circle*{.5}}\multiput(0,35)(2,2){6}{\circle*{.5}}
\multiput(0,37.5)(0,2.5){7}{\circle*{.5}}
\multiput(0,0)(2,2){5}{\circle*{.5}}
\put(0,0){\line(0,-1){2}}\put(-2,-9){$\scriptscriptstyle i=0$}
%
\put(0,33){\vector(0,-1){8}}
\end{picture}
&
\begin{picture}(85,20)(0,0)
\put(0,0){\line(1,0){60}}
{\thicklines
\put(0,0){\line(1,1){10}}
\put(10,10){\line(1,-1){10}}\put(20,0){\line(1,1){10}}
\put(30,10){\line(1,-1){10}}\put(40,0){\line(1,1){10}}
\put(50,10){\line(1,-1){10}}
}
\multiput(0,45)(2,2){5}{\circle*{.5}}\multiput(0,45)(2,-2){5}{\circle*{.5}}
\multiput(10,35)(2,2){5}{\circle*{.5}}\multiput(10,55)(2,-2){6}{\circle*{.5}}
\multiput(10,10)(2,2){10}{\circle*{.5}}
\multiput(30,30)(2,-2){10}{\circle*{.5}}
\put(0,0){\line(0,-1){2}}\put(-2,-9){$\scriptscriptstyle 0$}
\put(10,0){\line(0,-1){2}}\put(7,-9){$\scriptscriptstyle i=1$}
%
\put(10,33){\vector(0,-1){8}}
\end{picture}
&
\begin{picture}(85,20)(0,0)
\put(0,0){\line(1,0){60}}
\put(60,0){\line(0,1){20}}
{\thicklines
\put(0,0){\line(1,1){10}}
\put(10,10){\line(1,-1){10}}\put(20,0){\line(1,1){20}}
\put(40,20){\line(1,-1){10}}
\put(50,10){\line(1,1){10}}
}
\multiput(0,45)(2,2){5}{\circle*{.5}}\multiput(0,45)(2,-2){5}{\circle*{.5}}
\multiput(10,35)(2,2){5}{\circle*{.5}}\multiput(10,55)(2,-2){6}{\circle*{.5}}
\multiput(10,10)(2,2){15}{\circle*{.5}}
\multiput(40,40)(2,-2){5}{\circle*{.5}}
\multiput(50,30)(2,2){5}{\circle*{.5}}
\multiput(60,22.5)(0,2.5){8}{\circle*{.5}}
\put(0,0){\line(0,-1){3}}\put(-2,-9){$\scriptscriptstyle 0$}
\put(10,0){\line(0,-1){3}}\put(7,-9){$\scriptscriptstyle i=1$}
%
\put(10,33){\vector(0,-1){8}}
\end{picture}
&
\begin{picture}(75,20)(0,0)
\put(0,0){\line(1,0){50}}
\put(0,0){\line(0,1){10}}
\put(50,0){\line(0,1){20}}
{\thicklines
\put(0,10){\line(1,-1){10}}\put(10,0){\line(1,1){10}}
\put(20,10){\line(1,-1){10}}\put(30,0){\line(1,1){20}}
}
\multiput(10,45)(2,2){5}{\circle*{.5}}\multiput(10,45)(2,-2){5}{\circle*{.5}}
\multiput(20,35)(2,2){5}{\circle*{.5}}\multiput(20,55)(2,-2){6}{\circle*{.5}}
\multiput(0,30)(2,-2){10}{\circle*{.5}}
\multiput(20,10)(2,2){15}{\circle*{.5}}
\multiput(50,22.5)(0,2.5){8}{\circle*{.5}}
\multiput(0,12.5)(0,2.5){8}{\circle*{.5}}
\put(0,0){\line(0,-1){3}}\put(-2,-9){$\scriptscriptstyle 0$}
\put(20,0){\line(0,-1){3}}\put(17,-9){$\scriptscriptstyle i=2$}
%
\put(20,33){\vector(0,-1){8}}
\end{picture}
\\[5mm]
\mbox{a)~~~~~}&\mbox{b)~~~~~}&\mbox{c)~~~~~}&\mbox{d)~~~~~}&
\mbox{e)~~~~~}
\end{array}
$
\hfil
\caption{\footnotesize
The interface profile before event and the (half-)tiles
hitting the interface are drawn in dashed lines.
The interface profile after the event is drown in permanent line.
Pictures a) and b) illustrate absorption, respectively,
in a bulk and at the boundary of the interface.
Pictures  c) and d) are examples of  avalanches, respectively,
in the bulk (number of desorbed tiles $n_d=3$, size of a substrate
$L=6$) and near the boundary
($n_d=5$, $L=6$). Picture e) shows the total
avalanche ($n_d=L=5$).
}
\lb{absorb-desorb}
\end{figure}

For the model A the evolution rules described here were formulated
in Ref. \cite{GNPR2}. This stochastic process was named
{\em raise and peel model} (RPM) there. Models B and C are
versions of the RPM supplied with additional boundary terms.

\subsection{Stationary states: largest and smallest components
and normalization factors.}

From now on we will study
stationary states of the stochastic processes
(\ref{evol})--(\ref{3cases}),
i.e.  solutions of equation
\be
\lb{zerovec}
H_L |p_L{\cal i}\, =\, 0\, .
\ee
Note that the intensity property of matrix $H_L$
guarantees an existence of at least one nontrivial solution
of equation (\ref{zerovec}).

Below we  present results of a
numeric investigation of the
stationary states of three versions of RPM.
Calculations were carried out with the use of REDUCE
program for the system's  size  up to $L= 13/11/10$ for the models A/B/C,
respectively.
In all three cases the RPM has a unique stationary state.

Denote components of $|p_L{\cal i}$ (\ref{zerovec})
in the models A, B and C
as $p^{(a)}_L(w)$, $p^{(b)}_L(w)$
and $p^{(c)}_L(w)$, respectively.
Here argument $w$
labels in each case relevant interface configurations:
those are the sets of Dyck paths $\{w\}_{\footnotesize Dyck}$~ in the
model A,
Ballot paths $\{w\}_{\footnotesize Ballot}$~ in the model B
and Anchored Cross paths $\{w\}_{\footnotesize ACross}$~ in the model C.
Due to intensity property of the Hamiltonian
one always can choose null eigenvectors $|p_L{\cal i}$
in such a way that all their components are
nonnegative real (see \cite{ADR}), thus, making
the probabilistic interpretation consistent.
Moreover, it turns out that no one of coefficients
$p^{(*)}_L(w)$ vanishes.
So, we can normalize them
to be mutually primitive positive integers.
Denote their smallest and largest components as
\be
\lb{mM}
m_L^{(*)}:= \min_{\{w\}_{_*}}\{p_L^{(*)}(w)\}\, ,\qquad
M_L^{(*)}:= \max_{\{w\}_{_*}}\{p_L^{(*)}(w)\}\, .
\ee
It turns out that $m_L^{(a)}=m_L^{(b)}=1$,
but $m_L^{(c)}\neq 1$. The corresponding interface
configurations are shown on Fig.(\ref{Fig1a}).
\begin{figure}
\hfil
$
\begin{array}{cccccc}
\begin{picture}(80,50)(0,0)
\put(0,0){\line(1,0){60}}
{\thicklines
\put(0,0){\line(1,1){30}}\put(30,30){\line(1,-1){30}}
}
\multiput(10,10)(2,-2){5}{\circle*{.5}}
\multiput(20,0)(2,2){5}{\circle*{.5}}
\multiput(30,10)(2,-2){5}{\circle*{.5}}
\multiput(40,0)(2,2){5}{\circle*{.5}}
\end{picture}
&
\begin{picture}(75,50)(0,0)
\put(0,0){\line(1,0){50}}
\put(0,0){\line(0,1){10}}
{\thicklines
\put(0,10){\line(1,1){20}}\put(20,30){\line(1,-1){30}}
}
\multiput(0,10)(2,-2){5}{\circle*{.5}}
\multiput(10,0)(2,2){5}{\circle*{.5}}
\multiput(20,10)(2,-2){5}{\circle*{.5}}
\multiput(30,0)(2,2){5}{\circle*{.5}}
\multiput(40,10)(2,-2){5}{\circle*{.5}}
\end{picture}
&
\begin{picture}(75,50)(0,0)
\put(0,0){\line(0,1){10}}\put(0,0){\line(1,0){50}}
{\thicklines
\put(0,10){\line(1,-1){10}}\put(10,0){\line(1,1){20}}
\put(30,20){\line(1,-1){20}}
}
\multiput(20,10)(2,-2){5}{\circle*{.5}}
\multiput(30,0)(2,2){5}{\circle*{.5}}
\end{picture}
&
\begin{picture}(70,50)(0,0)
\put(0,0){\line(0,1){40}}\put(0,0){\line(1,0){40}}
{\thicklines
\put(0,40){\line(1,-1){40}}
}
\multiput(0,0)(2,2){5}{\circle*{.5}}
\multiput(10,10)(2,-2){5}{\circle*{.5}}
\multiput(20,00)(2,2){5}{\circle*{.5}}
\end{picture}
&
\begin{picture}(70,50)(0,0)
\put(40,0){\line(0,1){40}}\put(0,0){\line(1,0){40}}
{\thicklines
\put(0,0){\line(1,1){40}}
}
\multiput(10,10)(2,-2){5}{\circle*{.5}}
\multiput(20,0)(2,2){5}{\circle*{.5}}
\multiput(30,10)(2,-2){5}{\circle*{.5}}
\end{picture}
&
\begin{picture}(65,50)(0,0)
\put(30,0){\line(0,1){40}}\put(0,0){\line(1,0){30}}
\put(0,0){\line(0,1){10}}
{\thicklines
\put(0,10){\line(1,1){30}}
}
\multiput(0,10)(2,-2){5}{\circle*{.5}}
\multiput(10,0)(2,2){5}{\circle*{.5}}
\multiput(20,10)(2,-2){5}{\circle*{.5}}
\end{picture}
\\[5mm]
\mbox{a)~~~~~}&\mbox{b)~~~~~}&\mbox{c)~~~~~}&\mbox{d)~~~~~~}
&\mbox{e)~~~~~~}&\mbox{f)~~~~~~~}
\end{array}
$ \hfil \caption{\footnotesize For the model A the minimal possible
coefficient 1 in null eigenvector stands for pyramid type
configuration a) for $L$ even and for configurations b) and c) for
$L$ odd. For the model B half-pyramid configurations d) has coefficient
1 both for $L$ even and $L$ odd. For the model C the minimal coefficient
$m_L^{(c)}$ appears again for configurations d) and for e)/f) in
case of $L$ even/odd. Dashed lines show the substrate in each
case. } \lb{Fig1a}
\end{figure}
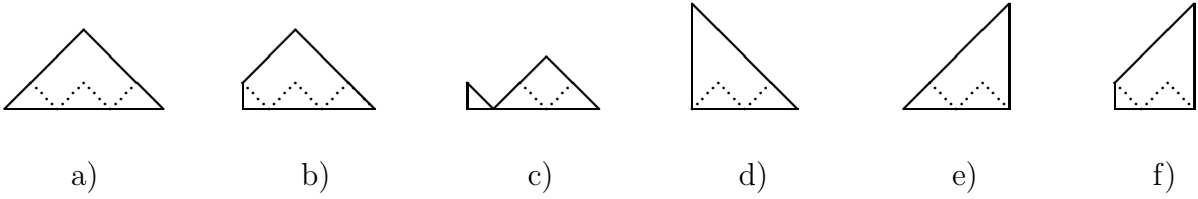

Denote total normalization factors of the stationary probability
distributions as
\be
\lb{S}
S_L^{(*)} := \sum_{\{w\}_{_*}} p^{(*)}_L(w)\, .
\ee

In the table below we collect values of $S_L^{(*)}$
and $m_L^{(c)}$ for $L\leq 11$
\ba
\lb{tab1}
\hspace{-2mm}
\begin{array}{|c|lllllllllll|}
\hline
L&\lefteqn{1} &\lefteqn{2} &3 &4 &5 &6 &7 &8 &9 &10 & 11\\
\hline
\scriptstyle S^{(a)}_L &
\lefteqn{\scriptstyle 1} &
\lefteqn{\scriptstyle 1} &
\scriptstyle 2 & \scriptstyle 3 &\scriptstyle 11 &
\scriptstyle 26 &\scriptstyle 170 & \scriptstyle 646 &
\scriptstyle 7429 &\scriptstyle 45885 &\scriptstyle 920460 \\[.5mm]
\scriptstyle S^{(b)}_L &
\lefteqn{\scriptstyle 1} &
\lefteqn{\scriptstyle 2} &
\scriptstyle 6 &\scriptstyle 33 &\scriptstyle 286~ &
\scriptstyle 4420~~ &\scriptstyle 109820~ &
\scriptstyle 4799134~~~ &\scriptstyle 340879665~~~~ &\scriptstyle 42235307100 &
\scriptstyle 8564558139000 \\[.5mm]
\scriptstyle S^{(c)}_L &
\lefteqn{\scriptstyle 2} &
\lefteqn{\scriptstyle 6} &
\lefteqn{\scriptstyle 66} &
\lefteqn{\scriptstyle 858} &
\lefteqn{\scriptstyle 48620} &
\lefteqn{\scriptstyle 1427660} &
\lefteqn{\scriptstyle 47991340} &
\lefteqn{\scriptstyle 11589908610} &
\lefteqn{\scriptstyle 13642004193300} &
\lefteqn{\scriptstyle 1139086232487000} & \\
\scriptstyle m^{(c)}_L &
\lefteqn{\scriptstyle 1} &
\lefteqn{\scriptstyle 1} &
\lefteqn{\scriptstyle 2} &
\lefteqn{\scriptstyle 3} &
\lefteqn{\scriptstyle 11} &
\lefteqn{\scriptstyle 13} &
\lefteqn{\scriptstyle 10} &
\lefteqn{\scriptstyle 34} &
\lefteqn{\scriptstyle 323} &
\lefteqn{\scriptstyle 133} &
\\
\hline
\end{array}
\ea
The table illustrates clearly the following three conjectures.
These conjectures  were  found also in \cite{BGN1,MNGB}.

\vspace{3mm}\noindent {\bf Conjecture~1.~}
{\em Let
$A^V_n$ (resp., $A^{VH}_n$) denote a  number of
vertically symmetric (resp., vertically
and horizontally symmetric)
alternating sign matrices of a size $n\times n$
\footnote{On enumeration of various symmetry classes of
alternating sign matrices see \cite{R,K}}.
Then
\be
\lb{sigma1}
S^{(a)}_{2p} = A^V_{2p+1} :=
(-3)^{p^2}\!\!\!\!\!\!\prod_{
\rule{0pt}{3mm}1\leq i\leq p\atop\rule{0pt}{3mm} 1\leq j\leq 2p+1}
\!\!{6i-3j+1\over 2i-j+2p+1}
\ ,\qquad S^{(b)}_L = A^{VH}_{2L+3}\ .
\ee}

\vspace{-4mm}\noindent{\bf Conjecture~2.~}{\em
Looking at numbers standing in down-up diagonals of the table one
observes  equalities
\ba
\lb{sigma2}
S^{(b)}_L &=& S^{(a)}_L \ S^{(a)}_{L+1}\ ,
\\[2mm]
\lb{mu}
S_L^{(c)} &=& m^{(c)}_L S^{(b)}_{L+1}\ .
\ea
}
The first equality (\ref{sigma2}) allows one to get
two expressions for $S^{(a)}_{2p-1}$~\footnote{
Note that $S^{(a)}_{2p-1}$ is just the number of
cyclically symmetric transpose complement plane partitions in a $(2p)^3$ box
(see, e.g., \cite{B}, p.199). It is usually denoted as $N_S(2p)$.}
\be
\lb{sigma3}
S^{(a)}_{2p-1}\ =\ {A^{VH}_{4p-1}/ A^V_{2p-1}}\ =\
{A^{VH}_{4p+1}/ A^V_{2p+1}} :=
\prod_{\rule{0pt}{3mm}0\leq i\leq p-1}
\!\!{(3i+1)(6i)!(2i)!\over (4i)!(4i+1)!}\, .
\ee
Their consistency, in turn, is based on relations conjectured in \cite{R}
\be
\lb{id}
A^{VH}_{4p+1}/ A^{VH}_{4p-1}\ =\
A^V_{2p+1}/ A^V_{2p-1}\ =\
{(3p-1){6p-3 \choose 2p-1}\over (4p-1){4p-2\choose 2p-1}}\, .
\ee

The second equality (\ref{mu}) relates quantities
$S^{(c)}_L$ and $m^{(c)}_L$. Looking at the table (\ref{tab1})
one can also assume that $m_L^{(c)}$ are divisors of $S^{(a)}_L$.
A formula for $m_L^{(c)}$ refining this observation is
guessed in \cite{MNGB} (see Eqs.(32), (33) there)

\be
\lb{mc}
m_L^{(c)} = \mbox{Numerator of}\; \left(
{S_L^{(a)}\over S_{L+2}^{(a)}}\right)\, .
\ee

\noindent\hspace*{2mm}
\parbox{163mm}{
{\bf Remark.~}
{\footnotesize
At this point one may propose another reasonable
normalization for the stationary state in the model C.
Denoting  components of the stationary state in this new normalization
as $\{{\widetilde p}^{(c)}_L\}$ one may fix normalization demanding that
\be
\lb{new-norm}
{\widetilde m}_L^{(c)}:= \min_{\{w\}_{_{ACross}}}\{{\widetilde p}_L^{(c)}(w)\}
= S_L^{(a)}\, ,
\ee
Certainly, the initial normalization is
a more economic one and thus it is better suited for calculations.
However, normalization (\ref{new-norm}) has an advantage in
interpreting results. E.g., formula for the total
normalization factors in this case looks as
\be
\lb{mu2}
{\widetilde S}_L^{(c)}\, =\, S_L^{(a)} S_{L+1}^{(b)}\, =\,
 S_L^{(b)} S_{L+2}^{(a)}\, =\,
 S_L^{(a)} S_{L+1}^{(a)}  S_{L+2}^{(a)}\, ,
\ee
which is much more in the spirit of eq.(\ref{sigma2})
then the formula (\ref{mu}).
We decide to keep initial `economic' normalization
throughout the text and to comment on the second
normalization when presenting results.
}}

\vspace{4mm}
The last conjecture of this section
describes  the largest  components
of the stationary states in the models A and B.
Analogous results for the model C are given in conjecture 13.

\vspace{2mm}\noindent{\bf Conjecture~3.~}{\em The maximal
coefficient $M^{(a)}_L$ appear in the set $\{p^{(a)}_L\}$ with
multiplicity 1 for $L$ even and $(L-1)/2$ for $L$ odd. The maximal
coefficient $M^{(b)}_L$ appear in the  set $\{p^{(b)}_L\}$ with
multiplicity 2 for $L$ even and 1 for $L$ odd. Their corresponding
interface configurations are shown on Fig.(\ref{Fig1}) and values
of the maximal coefficients are given by formulas \ba \lb{max} &&
M^{(a)}_L\  =\ S^{(a)}_{L-1}\ ,
\\[2mm]
\lb{max-b}
&& M^{(b)}_{L=2p}\  =\ (S^{(a)}_L)^2\ ,\quad
M^{(b)}_{L=2p+1}\  =\ S^{(a)}_{L-1} S^{(a)}_{L+1}\ .
\ea
}
\begin{figure}
\hfil
$
\begin{array}{cccc}
\begin{picture}(85,20)(0,0)
\put(0,0){\line(1,0){60}}
{\thicklines
\put(0,0){\line(1,1){10}}\put(10,10){\line(1,-1){10}}
\put(20,0){\line(1,1){10}}\put(30,10){\line(1,-1){10}}
\put(40,0){\line(1,1){10}}\put(50,10){\line(1,-1){10}}
}
\end{picture}
&
\begin{picture}(85,20)(0,0)
\put(0,0){\line(0,1){20}}\put(0,0){\line(1,0){60}}
{\thicklines
\put(0,20){\line(1,-1){20}}\put(20,0){\line(1,1){10}}
\put(30,10){\line(1,-1){10}}\put(40,0){\line(1,1){10}}
\put(50,10){\line(1,-1){10}}
}
\end{picture}
&
\begin{picture}(95,20)(0,0)
\put(0,0){\line(0,1){10}}\put(0,0){\line(1,0){70}}
{\thicklines
\put(0,10){\line(1,-1){10}}\put(10,0){\line(1,1){10}}
\put(20,10){\line(1,-1){10}}\put(30,0){\line(1,1){10}}
\put(40,10){\line(1,-1){10}}\put(50,0){\line(1,1){10}}
\put(60,10){\line(1,-1){10}}
}
\end{picture}
&
\begin{picture}(120,20)(0,0)
\put(0,-4){\line(0,1){14}}\put(0,0){\line(1,0){78}}
\put(81,0){\dots}\put(96,0){\line(1,0){29}}
\put(50,0){\line(0,-1){4}}\put(-2,-13){$\scriptstyle 0$}
\put(48,-13){$\scriptstyle k, \;\; k=1,3,5,\dots ,L.$}
\put(125,0){\line(0,-1){4}}\put(123,-13){$\scriptstyle L$}
{\thicklines
\put(0,10){\line(1,1){10}}\put(10,20){\line(1,-1){10}}
\put(20,10){\line(1,1){10}}
\put(30,20){\line(1,-1){20}}\put(50,0){\line(1,1){10}}
\put(60,10){\line(1,-1){10}}\put(70,0){\line(1,1){10}}
\put(95,10){\line(1,-1){10}}\put(105,0){\line(1,1){10}}
\put(115,10){\line(1,-1){10}}
}
\end{picture}
\\[5mm]
\mbox{a)~~~~~}&\mbox{b)~~~~~}&\mbox{c)~~~~~}&\mbox{d)}
\end{array}
$ \hfil \caption{\footnotesize In the model A the maximal coefficient
in null eigenvector stands for configuration a) for $L$ even and
for configurations d) for $L$ odd. In the model B configurations a) and
b) enter the null eigenvector with the maximal coefficient if $L$
is even and configuration c) corresponds to maximal coefficient if
$L$ is odd. } \lb{Fig1}
\end{figure}
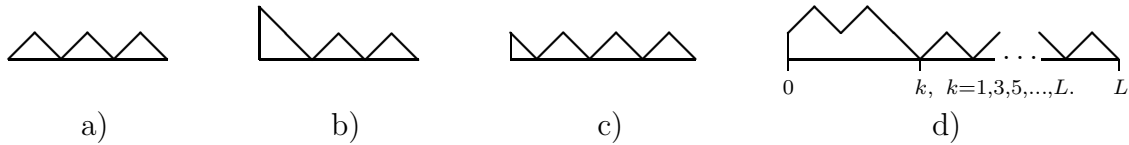

\subsection{Model A: detailed distributions and Pascal's hexagon relation}

Let us first introduce  more notation.

An integer point $i$ of an interface configuration $\{h_k\}_{k=0,\dots ,L}$
such that $0<i<L$ for the model A; $0\leq i<L$ for the model B; or, $0\leq i\leq L$
for the model C, is called an {\em $N$-contact}  if
$h_i=N$ and $h_{i\pm 1}=N+1$.
The contact points are  local minima of the interface
at which it can  absorb (half-)tiles.
For instance, the interface shown on picture d) of Fig.(\ref{Fig1})
has only one 1-contact $i=2$ in case A, but it has
two 1-contacts $i=0,2$ in cases B, or C.
In all cases it has 0-contacts
at points $k,k+2,\dots L-2$  and additionally
it has 0-contact at point $L$ in case C.

For either of the families of Dyck, Ballot, or Anchored Cross
paths we  denote $\{w\}^N_{_*}$ (here $*$ stands for an
appropriate family name) the subsets of all interface
configurations which have no $p$-contacts with $p<N$. These are
the configurations whose global minimum (excluding points $0$ and
$L$ for the model A, or point $L$ for the model B) is higher or equal to $N$.
Obviously,  one has $\{ w\}^0_{_*} \equiv \{w\}_{_*}$ and
$\{w\}^{N+1}_{_*}\subset\{w\}^N_{_*}$. We shall stress that there
is no strict correlation between the subsets $\{w\}^N_{_*}$
corresponding to different sets of paths. For example, for any $N$
$$
\{w\}^N_{Dyck}\subset
\{w\}_{Ballot}^{k}\, ,\qquad
\{w\}^N_{Dyck}\cap
\{w\}_{Ballot}^{k+1}\, =\, \emptyset\, ,
$$
where $k=0/1$ for $L$ even/odd. This is because
the left boundary point $h_0=0, \mbox{~or~} 1$
of a Dyck path is treated as 0-, or 1-contact point in
a family of Ballot paths.

Considering the subsets $\{w\}_{*}^N$ proves to be useful for
 analysis of the raise and peel model in cases A and B. Here
label $N$ spans integers from $0$ to $[(L-1)/2]$ in  case A
and from $0$ to $(L-1)$ in case B.\footnote{
By contrast, for the model C label $N$ can take only two values --- 0 and 1.
}
The subset with highest possible $N$
in both models A and B contains  only element
which is drawn on Fig.(\ref{Fig1a}),
pictures a)/b) and d), respectively.
As it was mentioned before,
this element enters the stationary state with a minimal
coefficient 1.
In the sequel we determine detailed distributions
$S_{L,N}^{(*)}$
and detailed maxima $M_{L,N}^{(*)}$
for each of the subsets $\{w\}_{_*}^N$
\be
\lb{MS-detailed}
S^{(*)}_{L,N}:= \sum_{\rule{0pt}{2.5mm}\{w\}_{_*}^N} p^{(*)}_L(w)\, , \qquad
M_{L,N}^{(*)}:= \max_{\rule{0pt}{3mm}\{w\}_{_*}^N}\{p_L^{(*)}(w)\}\, .
\ee

In this section we concentrate on studying the model A.
The table below contains values of $S_{L,N}^{(a)}$
for $L$ up to 13.
\ba
\lb{tab4}
\hspace{-4mm}
\begin{array}{|l|llllllllllllll|}
\hline
L\backslash N&
\multicolumn{1}{c|}{\scriptstyle \,\, -1~\,}
\lefteqn{
\begin{picture}(0,0)
\linethickness{.15pt}
\put(-.2,-4){\line(0,-1){162}}
\put(-.2,-179){\line(0,-1){14}}
\end{picture}} &
\phantom{\scriptstyle 0~~~}
\lefteqn{\hspace{1.4mm}\scriptstyle 0
\begin{picture}(0,0)
\linethickness{.15pt}
\put(23.3,-4){\line(0,-1){162}}
\put(23.3,-179){\line(0,-1){14}}
\end{picture}} &
\multicolumn{1}{l|}{\phantom{\scriptstyle 1/2~}} &
\phantom{\scriptstyle 1~~~}
\lefteqn{\hspace{1.4mm}\scriptstyle 1
\begin{picture}(0,0)
\linethickness{.15pt}
\put(23.3,-4){\line(0,-1){162}}
\put(23.3,-179){\line(0,-1){14}}
\end{picture}} &
\multicolumn{1}{l|}{\phantom{\scriptstyle 1/2~}} &
\phantom{\scriptstyle 2~~~\,}
\lefteqn{\hspace{1.4mm}\scriptstyle 2
\begin{picture}(0,0)
\linethickness{.15pt}
\put(23.3,-4){\line(0,-1){189}}
\end{picture}} &
\multicolumn{1}{l|}{\phantom{\scriptstyle 1/2~}} &
\phantom{\scriptstyle 3~~~}
\lefteqn{\hspace{1.4mm}\scriptstyle 3
\begin{picture}(0,0)
\linethickness{.15pt}
\put(23.3,-4){\line(0,-1){189}}
\end{picture}} &
\multicolumn{1}{l|}{\phantom{\scriptstyle 1/2~}} &
\phantom{\scriptstyle 4~~~}
\lefteqn{\hspace{1.4mm}\scriptstyle 4
\begin{picture}(0,0)
\linethickness{.15pt}
\put(23.3,-4){\line(0,-1){189}}
\end{picture}} &
\multicolumn{1}{l|}{\phantom{\scriptstyle 1/2~}} &
\phantom{\scriptstyle 5~~~}
\lefteqn{\hspace{1.4mm}\scriptstyle 5
\begin{picture}(0,0)
\linethickness{.15pt}
\put(23.3,-4){\line(0,-1){189}}
\end{picture}} &
\multicolumn{1}{l|}{\phantom{\scriptstyle 1/2~}} &
\phantom{\scriptstyle 6~~~}
\lefteqn{\hspace{-3.5mm}\scriptstyle 6}

\\
\hline
\scriptstyle 1~~~ &  &\lefteqn{\scriptstyle 1}&&&&&&&&&&&&
\\
\scriptstyle 2 &  \lefteqn{\scriptstyle 1} &&
\lefteqn{\scriptstyle 1}&&&&&&&&&&&
\\
\scriptstyle 3 &&  \lefteqn{\scriptstyle 2} && \lefteqn{\scriptstyle 1}
&&&&&&&&&&
\\
\scriptstyle 4 &\lefteqn{\scriptstyle 3}&& \lefteqn{\scriptstyle 3} &&
\lefteqn{\scriptstyle 1} &&&&&&&&&
\\
\scriptstyle 5 && \lefteqn{\scriptstyle 11} && \lefteqn{\scriptstyle 4} &&
\lefteqn{\scriptstyle 1} &&&&&&&&
\\
\scriptstyle 6 &\lefteqn{\scriptstyle 26}&& \lefteqn{\scriptstyle 26} &&
\lefteqn{\scriptstyle 5} && \lefteqn{\scriptstyle 1} &&&&&&&
\\
\scriptstyle 7 && \lefteqn{\scriptstyle 170} && \lefteqn{\scriptstyle 50}&&
\lefteqn{\scriptstyle 6} && \lefteqn{\scriptstyle 1} &&&&&&
\\
\scriptstyle 8 &\lefteqn{\scriptstyle 646}&& \lefteqn{\scriptstyle 646} &&
\lefteqn{\scriptstyle 85
\begin{picture}(0,0)
\thinlines
\put(2,2){\line(1,0){40}}
\end{picture}}
&& \lefteqn{\scriptstyle 7} && \lefteqn{\scriptstyle 1}
&&&&&
\\
\scriptstyle 9 && \lefteqn{\scriptstyle 7429} && \lefteqn{\scriptstyle 1862
\begin{picture}(0,0)
\thinlines
\put(-3,8){\line(2,1){10}}
\put(-3,-4){\line(2,-1){10}}
\end{picture}} &&
\lefteqn{\scriptstyle 133} && \lefteqn{\begin{picture}(0,0)
\thinlines
\put(-2,8){\line(-2,1){14}}
\put(-2,-4){\line(-2,-1){10}}
\end{picture}\scriptstyle 8} &&
\lefteqn{\scriptstyle 1} &&&&
\\
\scriptstyle 10 &\lefteqn{\scriptstyle 45885}&& \lefteqn{\scriptstyle 45885}
 && \lefteqn{\scriptstyle 4508
\begin{picture}(0,0)
\thinlines
\put(2,2){\line(1,0){32}}
\end{picture}} &&
\lefteqn{\scriptstyle 196} && \lefteqn{\scriptstyle 9} &&
\lefteqn{\scriptstyle 1 }&&&
\\
\scriptstyle 11 && \lefteqn{\scriptstyle 920460} && \lefteqn{\scriptstyle 202860}
&& \lefteqn{\scriptstyle 9660} && \lefteqn{\scriptstyle 276} &&
\lefteqn{\scriptstyle 10} && \lefteqn{\scriptstyle 1}&&
\\
\scriptstyle 12 &\lefteqn{\scriptstyle 9304650}&&
\lefteqn{\scriptstyle 9304650} &&
\lefteqn{\scriptstyle 720360} &&
\lefteqn{\scriptstyle 18900}&& \lefteqn{\scriptstyle 375} &&
\lefteqn{\scriptstyle 11} && \lefteqn{\scriptstyle 1}&
\\
\scriptstyle 13 && \lefteqn{\scriptstyle 323801820} &&
\lefteqn{\scriptstyle 64080720} &&
\lefteqn{\scriptstyle 2184570} && \lefteqn{\scriptstyle 34452} &&
\lefteqn{\scriptstyle 495} && \lefteqn{\scriptstyle 12} &&\lefteqn{\scriptstyle 1}
\\
\hline
\end{array}
\ea
\smallskip

\noindent
For later convenience
we add to the table a column for $N=-1$ setting
$S^{(a)}_{L=2p,N=-1}:=S^{(a)}_{L=2p,N=0}$.
We also align data
corresponding to even, or odd values of $L$
leftwards, or rightwards in the columns, correspondingly.
The numbers $S^{(a)}_{L,N=0}\equiv S^{(a)}_L$
were already listed in
table (\ref{tab1}).

It is helpful to look at the
up-down diagonals in this table. The rightmost diagonal contains units only.
the next one contains numbers $(L-1)$.
For numbers staying in
diagonals from third to seventh one finds expressions
\ba
\lb{3diag}
&&\hspace{-14mm}
{\scriptstyle(L-2)(L-3)\cdot(2L+1)\over\scriptstyle 2\cdot 3}\, ,
\\[2mm]
\lb{4diag}
&&\hspace{-14mm}
{\scriptstyle(L-2)(L-3)(L-4)(L-5)\cdot(2L+1)(2L+3)\over\scriptstyle
 2^2\cdot 3^2\cdot 5}\, ,
\\[2mm]
\lb{5diag}
&&\hspace{-14mm}
{\scriptstyle(L-3)(L-4)^2 (L-5)(L-6)(L-7)
\cdot(2L-1)(2L+1)(2L+3)(2L+5)\over\scriptstyle
2^4\cdot 3^3\cdot 5^2\cdot 7}\, ,
\\[2mm]
\lb{6diag}
&&\hspace{-14mm}
{\scriptstyle (L-3)(L-4)(L-5)^2(L-6)^2(L-7)(L-8)(L-9)
\cdot(2L-1)(2L+1)^2 (2L+3)(2L+5)(2L+7)\over\scriptstyle
2^6\cdot 3^4\cdot 5^3\cdot 7^2\cdot 9}\, ,
\\[2mm]
\lb{7diag}
&&\hspace{-14mm}
{\scriptstyle (L-4)(L-5)^2(L-6)^2(L-7)^2(L-8)^2(L-9)(L-10)(L-11) \cdot
(2L-3)(2L-1)(2L+1)^2 (2L+3)^2(2L+5)(2L+7)(2L+9)\over\scriptstyle
2^9\cdot 3^5\cdot 5^4\cdot 7^3\cdot 9^2\cdot 11}\, .
\ea
With these data one can guess general formula for $S^{(a)}_{L,N}$.

\vspace{3mm}\noindent {\bf Conjecture~4 (Model A: detailed
distributions).~} {\em Denote $n:=[{L-1\over 2}]-N$. Integer~ $n$
starts from 0 and labels leftwards the up-down diagonals in the
table (\ref{tab4}). One has
\ba
\lb{Sa}
\hspace*{-5mm}
S^{(a)}_{L,N}\, =\, 2^{^{-[n^2/ 4]}} \prod_{p=1}^{n}{1\over
(2p-1)!!}\, \prod_{p=0}^{[{n-1\over 3}]}{\bigl(L-[{n+p\over
2}]-p-1\bigr)!\over \bigl(L-2n+3p\bigr)!}\,
\prod_{p=0}^{[{n-2\over 3}]}{\bigl(2L+2n-6p-3\bigr)!!\over
\bigl(2L-2[{n+p\over 2}]+4p+1\bigr)!! }
\ea
}

The same sequence was guessed in \cite{MNGB}~\footnote{
For a particular case
$L=2p$, $N=1$ distributions $S^{(a)}_{2p,1}$
coincide with $P_p(1)$, where $P_p(k)$ is the unnormalized
probability to have $k$ clusters in the stationary state
(see \cite{GNPR2}).
An expression for $P_p(k)$
is given in Conjecture 3 in \cite{G}.}.
Our notation is related to that used in \cite{MNGB}
as $S^{(a)}_{L,N=[(L-1)/2]-n}\equiv R(n+1,L+1)$~
(see eq.(17) there).

\medskip
As it is explained, e.g.,  in \cite{G,MNGB} the numbers
given by formula (\ref{Sa})
appear in  counting elements of certain families of vertically
symmetric alternating sign matrices.
It is not however clear how one can characterize these
families of ASMs.
The table (\ref{tab4}) suggests another  combinatorial
interpretation of integers $S^{(a)}_{L,N}$.

\vspace{3mm}\noindent {\bf Pascal's hexagon relations.~}
{\em
The numbers in sequence $S^{(a)}_{L,N}$ (\ref{Sa}) satisfy equalities
\ba
\lb{hexagon1}
S^{(a)}_{L-1,N} S^{(a)}_{L+1,N+1} + S^{(a)}_{L,N-1} S^{(a)}_{L,N+1}
&=&
S^{(a)}_{L-1,N+1} S^{(a)}_{L+1,N}\, , \quad
\mbox{if $L$ is even}\, ,
\\[2mm]
\lb{hexagon2}
S^{(a)}_{L-1,N} S^{(a)}_{L+1,N+1} + S^{(a)}_{L,N} S^{(a)}_{L,N+2}
&=&
S^{(a)}_{L-1,N+1} S^{(a)}_{L+1,N}\, , \quad
\mbox{if $L$ is odd}\, .
\ea
Substituting label $N$
by \mbox{$n:=([{L-1\over 2}]-N$)} in the notation $S^{(a)}_{L,N}$
one can uniformly write relations (\ref{hexagon1}), (\ref{hexagon2})
as
\be
\lb{uniform}
S^{(a)}_{L-1,n} S^{(a)}_{L+1,n} + S^{(a)}_{L,n-1} S^{(a)}_{L,n+1}\ =\
S^{(a)}_{L-1,n-1} S^{(a)}_{L+1,n+1}\, .
\ee
}

In table (\ref{tab4}) numbers participating in
eq.(\ref{uniform}) form a hexagonal structure.
One such hexagon is drawn in the table.
The components
of each square monomial in (\ref{uniform}) occupy opposite vertices of
the hexagon. Relation (\ref{uniform}) thus looks as
a sophisticated variant of Pascal's triangle relation,
wherefrom our notation follows.

Like in Pascal's triangle case
relations (\ref{uniform}) can be used to reconstruct all  numbers
$S^{(a)}_{L,n}$ provided some, say boundary, part
of them are fixed.
For certain boundary data the numbers $S^{(a)}_{L,n}$ turn out to be integers
and one recognizes among them
the numbers of vertically symmetric and half turn symmetric
alternating sign matrices, $A^V_{2n+1}$ and $A^{HT}_n$, whereas
the total number of ASMs $A_n$, and the number of vertically and horizontally
symmetric ASMs $A^{VH}_n$ appear in combinations.
A brief discussion of this subject is presented in the Appendix.

\smallskip
We finish the section with observations of detailed maxima and
of symmetry properties of the stationary distribution.

\vspace{2mm}\noindent{\bf Conjecture~5 (Model A: detailed
maxima).~}{\em In the stationary state of the model A maximal
coefficient $M^{(a)}_{L,N}$ in the  subset $\{w\}^N_{Dyck}$ stands
for configuration with a maximal possible number $([(L-1)/2]-N)$
of $N$-contacts (see explanations on Fig.(\ref{det-max})). It is
given by formula \be \lb{max2} M^{(a)}_{L,N}\  =\
S^{(a)}_{L-1,N-\epsilon(L)}\ , \ee where
$\epsilon(L):={1-(-1)^L\over 2}$ is a parity
function.}\,\footnote{ Here and everywhere below in the main text
we use  definition of $S^{(a)}_{L,N}$ as it is given in
eq.(\ref{Sa}). This notation is natural for the stochastic models
we are treating. Different convention about  the second index in
$S^{(a)}_{L,n}$ is used in eq.(\ref{uniform}) and in the Appendix.
It is more suitable from a mathematical viewpoint. }
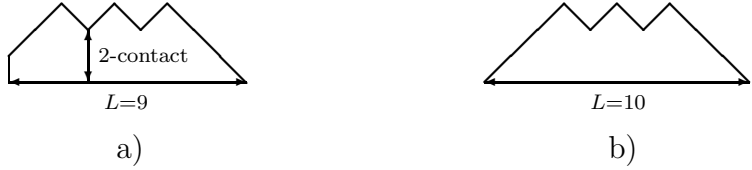
\begin{figure}
\hfil
$
\begin{array}{cc}
\begin{picture}(170,20)(0,0)
\put(0,0){\vector(1,0){90}} \put(90,0){\vector(-1,0){90}}
\put(30,0){\vector(0,1){20}}\put(30,20){\vector(0,-1){20}}
\put(34,8){\scriptsize 2-contact}
 \put(0,0){\line(0,1){10}}
{\thicklines
\put(0,10){\line(1,1){20}}\put(20,30){\line(1,-1){10}}
\put(30,20){\line(1,1){10}}\put(40,30){\line(1,-1){10}}
\put(50,20){\line(1,1){10}}\put(60,30){\line(1,-1){30}}
\put(36,-10){$\scriptstyle L=9$}}
\end{picture}
&
\begin{picture}(140,20)(0,0)
\put(0,0){\vector(1,0){100}} \put(100,0){\vector(-1,0){100}}
{\thicklines \put(0,0){\line(1,1){30}}\put(30,30){\line(1,-1){10}}
\put(40,20){\line(1,1){10}}\put(50,30){\line(1,-1){10}}
\put(60,20){\line(1,1){10}}\put(70,30){\line(1,-1){30}}
\put(40,-10){$\scriptstyle L=10$}}
\end{picture}
\\[5mm]
\mbox{a)~~~~~~~~~~~~~~~~~~~~}&\mbox{b)~~~~~~~~~}
\end{array}
$ \hfil \caption{\footnotesize In the model A configurations a)/b) of a
size $L=9/10$ have no 0- and 1-contacts. Thus, they belong to
subset $\{w\}^{N=2}_{Dyck}$. In this subset among all the
configurations of a fixed size $L=9/10$ they have maximal possible
number $n=2$ of 2-contacts and in accordance with the statement of
Conjecture 5 they enter the stationary state with a maximal
coefficients $M^{(a)}_{L=9/10,N=2}$. } \lb{det-max}
\end{figure}

\vspace{2mm} The relation (\ref{max2}) was also observed in
\cite{MNGB}. Note that in case $N=0$ and $L$ odd we use in
eq.(\ref{max2}) an extension of sequence $S^{(a)}_{L,N}$ to case
$N=-1$ that we made in table (\ref{tab4}). Note also that equality
$M^{(a)}_{L=2p+1,N=0}=M^{(a)}_{L=2p+1,N=1}$ following from
(\ref{max2}) agrees with the statements made in Conjecture 3 (see
Fig.(\ref{Fig1}), picture d), cases $k=1$ and $k=L$).

\vspace{2mm}\noindent{\bf Conjecture~6.~}{\em
In the model A   for any
configuration $\{h_i\}$
there exists  a configuration $\{h'_i\}$ which appears
in the stationary state with the same coefficient.
The definition of $\{h'_i\}$ is following

\noindent
-- for $L$ even, $h'_i := h_{L-i}$;

\noindent
-- for $L$ odd, let $i_0$ be coordinate of the leftmost 0-contact
point in configuration $\{h_i\}$, or  $i_0=L$ if there is no 0-contacts in
$\{h_i\}$. Then, $h'_i:= h_{L-i}+1$ for all $i\leq L-i_0$, and
$h'_i:= h_{L-i}-1$ for all $i> L-i_0$.
}

\vspace{2mm} For $L$ even the statement of Conjecture 6 is a
direct consequence of the left-right symmetry of the Hamiltonian
and the configuration space. For $L$ odd the space of the states
is no more left-right symmetric, and the symmetry is not an
obvious one. Typical pair of symmetric configurations is drawn
below (pieces of paths A/B and A'/B' are left-right symmetric).

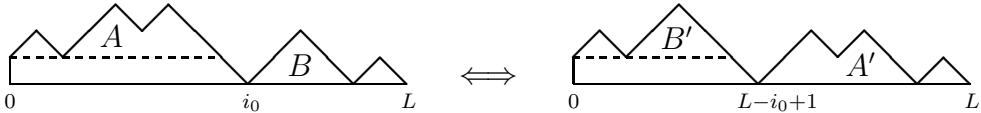
\hfil
$
\begin{array}{ccc}
\begin{picture}(160,32)(0,0)
\put(34,14){$\scriptsize A$} \put(105,3){$\scriptsize B$}
\put(0,0){\line(1,0){150}} \put(0,0){\line(0,1){10}}
\multiput(0,10)(5,0){16}{\line(1,0){2.5}} {\thicklines
\put(0,10){\line(1,1){10}}\put(10,20){\line(1,-1){10}}
\put(20,10){\line(1,1){20}}\put(40,30){\line(1,-1){10}}
\put(50,20){\line(1,1){10}}\put(60,30){\line(1,-1){30}}
\put(90,0){\line(1,1){20}}\put(110,20){\line(1,-1){20}}
\put(130,0){\line(1,1){10}}\put(140,10){\line(1,-1){10}} }
\put(-2,-9){$\scriptstyle 0$}\put(88,-9){$\scriptstyle
i_0$}\put(148,-9){$\scriptstyle L$}
\end{picture}
& \Longleftrightarrow ~~ &
\begin{picture}(160,32)(0,0)
\put(33,13){$\scriptsize B'$} \put(103,3){$\scriptsize A'$}
\put(0,0){\line(1,0){150}} \put(0,0){\line(0,1){10}}
\multiput(0,10)(5,0){12}{\line(1,0){2.5}}{\thicklines
\put(0,10){\line(1,1){10}}\put(10,20){\line(1,-1){10}}
\put(20,10){\line(1,1){20}}\put(40,30){\line(1,-1){30}}
\put(70,0){\line(1,1){20}}\put(90,20){\line(1,-1){10}}
\put(100,10){\line(1,1){10}}\put(110,20){\line(1,-1){20}}
\put(130,0){\line(1,1){10}}\put(140,10){\line(1,-1){10}}
\put(62,-9){$\scriptstyle L-i_0+1$} \put(-2,-9){$\scriptstyle
0$}\put(148,-9){$\scriptstyle L$}}
\end{picture}
\end{array}
$
\hfil

\subsection{Model B: left orbits and factorizations to model A}

Again, we start with a few definitions.

Consider a configuration $w=\{h_k\}_{k=0,\dots ,L}$ from the set of
Anchored Cross paths. For any $N$-contact point $i$ of the configuration
$w$
we construct new configurations $w_l(i)$ and $w_r(i)$
(here value of $N$ is irrelevant)
\ba
\lb{w'}
w_{l}(i) = \{ h'_k\} : && h'_k :=\left\{
\begin{array}{rcl}
h_k+2 & \mbox{for}& 0\leq k\leq i\, ,
\\
h_k &\mbox{for}& i<k\leq L\, ,
\end{array}
\right.
\\[2mm]
\lb{w''}
w_{r}(i) = \{h''_k \} : &&h''_k :=\left\{
\begin{array}{rcl}
h_k & \mbox{for}& 0\leq k< i\, ,
\\
h_k+2 &\mbox{for}& i\leq k\leq L\, ,
\end{array}
\right.
\ea
It may happen that either one or both configurations $w_l(i)$ and $w_r(i)$
defined by eqs.(\ref{w'}), (\ref{w''}) do not satisfy
condition c) of the definition (\ref{hi}) of Anchored Cross paths.
Such  configurations $w_l(i)/w_r(i)$ should
undergo total avalanche (see the end of Sec. 3.1),
i.e., they are to be redefined as
\be
\lb{w-add}
\{h_{_k}\!\!{'}^/{''}\}_{k=0,\dots ,L}\longrightarrow
\{h_{_k}\!\!{'}^/{''}-2\}_{k=0,\dots ,L}\, ,
\ee
Thus defined  configurations $w_l(i)$ and $w_r(i)$
are called, respectively,
{\em left and right coverings of the configuration $w$
at the point $i$}. Below, using the notion of left and right coverings
we give an iterative description of left and right orbits
of  the configuration $w$.

Let $L(w,N)/R(w,N)$ denote a set of all $N$-contacts of $w$ which
do not occupy right/left boundary point and which are placed to
the left/right of any $k$-contact with $k<N$. First, we construct
left/right coverings $w_l(i)/w_r(j)$ of $w$ at all points $i\in
L(w,N)/j\in R(w,N)$. Note that for all coverings $w_l(i)/w_r(j)$
the sets $L(w_l(i),N)/R(w_r(j),N)$ lie strictly inside the set
$L(w,N)/R(w,N)$. Next, we construct left/right coverings for all
obtained at a first step configurations $w_l(i)/w_r(j)$ at all
points from their corresponding sets $L(w_l(i),N)/R(w_r(j),N)$. We
repeat the procedure until at some step of iteration there will be
no more $N$-contacts available for generation of new coverings.
The set of all thus obtained left/right coverings together with
$w$ itself is called {\em left/right $N$-orbit of the
configuration $w$} and is denoted as ${\cal O}_l(w,N)/ {\cal
O}_r(w,N)$; $w$ is called {\em generating element} of the orbit.
The process of iterative construction of left and right orbits is
illustrated on Fig.6, p.\pageref{FigAA}  and Fig.7,
p.\pageref{FigB}, respectively. One can see that the numbers of
elements in the orbits ${\cal O}_l(w,N)/ {\cal O}_r(w,N)$ and in
the sets $L(w,N)/R(w,N)$ are related as \be \lb{rela} \# {\cal
O}_l(w,N)\, =\, 2^{\# L(w,N)}\, , \qquad \# {\cal O}_r(w,N)\, =\,
2^{\# R(w,N)}\, . \ee

An $N$-orbit is called {\em maximal} if it is not a subset of some
bigger $N$-orbit.
\medskip

For the family of Ballot paths only the procedure of left covering
makes sense and, hence, only the left orbits  are  defined. In the
rest of this section we apply notion of the left orbits to
establish detailed relations for the stationary states in case B
raise and peel model. The maximal 0- and 1-orbits will be
especially important for us. The reasons are following.
\medskip

The family of Ballot paths splits uniquely into collection of all
(mutually nonintersecting) maximal left 0-/1-orbits.

The maximal left 0-orbits of length $L$
are generated by  elements
\be
\lb{ah1}
w=\{h_k\}_{k=0,\dots,L}:\quad
h_0 = \left\{
\begin{array}{rl}
0 & \mbox{for $L$ even},
\\
1 & \mbox{for $L$ odd}.
\end{array}
\right.
\ee

The maximal left 1-orbits of length $L$
are generated by elements
\be
\lb{ah2}
w=\{h_k\}_{k=0,\dots,L}:\quad
h_0 = \left\{
\begin{array}{rl}
\mbox{either~} 0, \mbox{~or~} 2 & \mbox{for $L$ even},
\\
1 & \mbox{for $L$ odd}.
\end{array}
\right.
\ee
\medskip

We further observe that sums of the stationary coefficients
over maximal left 0- and 1-orbits
in case B are related to certain stationary coefficients in case A.
These relations generalizing formula (\ref{sigma2}) are presented in

\vspace{2mm}\noindent{\bf Conjecture~7.~}{\em
Take any Dyck paths $u$ and $v$ of  the
lengths $L$ and $(L+1)$, respectively.
Denote $^*\! v$ a Ballot path of the length $L$ which is a (left) reduction
of the Dyck path $v$, i.e.,
$$
h_k({^*}\! v) = h_{k+1}(v)\, , \mbox{~~for all~~} k=0,\dots ,L\, .
$$
With these notations one has
\ba
\lb{7a}
\hspace{-15mm}
\mbox{for $L$ even:}&&
\displaystyle \sum_{\nu\in{\cal O}_l(u,0)}p_L^{(b)}(\nu)\, =\, S^{(a)}_{L+1}\,
p^{(a)}_L(u)\, ,
\quad
\sum_{\nu\in{\cal O}_l({^*}\! v,1)}p_L^{(b)}(\nu)\, =\, S^{(a)}_L\,
p^{(a)}_{L+1}(v)\, ,
\\[5mm]
\lb{7b}
\hspace{-15mm}
\mbox{for $L$ odd:}&&
\displaystyle \sum_{\nu\in{\cal O}_l({^*}\! v,0)}p_L^{(b)}(\nu)\, =\, S^{(a)}_L\,
p^{(a)}_{L+1}(v)\, ,
\quad
\displaystyle \sum_{\nu\in{\cal O}_l(u,1)}p_L^{(b)}(\nu)\, =\, S^{(a)}_{L+1}\,
p^{(a)}_L(u)\, .
\ea
Here summation is taken over all elements of the left 0-, or 1-orbits
generated by $u$, or $^*\! v$ (note that configurations
$u$ and $^*\! v$  span all
generating elements described in (\ref{ah1}), (\ref{ah2})).
}
\medskip

Combining the last conjecture  with conjectures 4--6 from the previous section
one can get explicit values for sums over certain
0- and 1-orbits. These results and similar formulas for
higher orbits are presented in the next

\vspace{2mm}\noindent{\bf Conjecture~8.~}{\em
Let us denote $W(h_0,s) =\{h_k\}_{k=0,\dots ,L}$
a Ballot path of a size $L$
such that for a given left boundary height $h_0$
it has maximal possible number of $s$-contact points.
We are interested in configurations where $s$ take values between
$max\{0,(h_0-1)\}$ and $({L+h_0\over 2}-1)$ (see Fig.(\ref{generators})).

One can guess explicit expressions for sums of the coefficients
$p_L^{(b)}$ over left $h_0$- and $(h_0-1)$-orbits generated by
elements $W(h_0,s)$. For even size $L$ parameter $h_0$ takes on
even values from 0 to $L$ and one has relations
\ba
\lb{p-sums-a}
\begin{array}{rrcl}
\mbox{ $(h_0=2m)$-orbits:}\hspace{10mm}&\displaystyle
\sum_{\nu\in{\cal O}_l(W(h_0=2m,s),2m)}p_L^{(b)}(\nu)& =&
S^{(a)}_{L+1,m}\,
S^{(a)}_{L-1,s-m}\, ,
\\[6mm]
\mbox{ $(h_0-1=2m-1)$-orbits:}\hspace{10mm}&\displaystyle
\sum_{\nu\in{\cal O}_l(W(h_0=2m,s),2m-1)}p_L^{(b)}(\nu)& =&
S^{(a)}_{L,m-1}\,
S^{(a)}_{L,s-m}\, .
\end{array}
\ea
The last formula is applicable in case $m=0$, i.e., for (-1)-orbits.
In this case all the (-1)-orbits are treated as singlets containing their
generating elements only.

For odd size $L$ parameter $h_0$ takes on odd values from 1 to $L$ and one
has relations
\ba
\lb{p-sums-b}
\begin{array}{rrcl}
\mbox{ $(h_0-1=2m)$-orbits:}\hspace{10mm}&\displaystyle
\sum_{\nu\in{\cal O}_l(W(h_0=2m+1,s),2m)}p_L^{(b)}(\nu)& =&
S^{(a)}_{L,m}\,
S^{(a)}_{L,s-m}\, ,
\\[6mm]
\mbox{ $(h_0=2m+1)$-orbits:}\hspace{10mm}&\displaystyle
\sum_{\nu\in{\cal O}_l(W(h_0=2m+1,s),2m+1)}p_L^{(b)}(\nu)& =&
S^{(a)}_{L+1,m}\,
S^{(a)}_{L-1,s-m-1}\, .
\end{array}
\ea
}
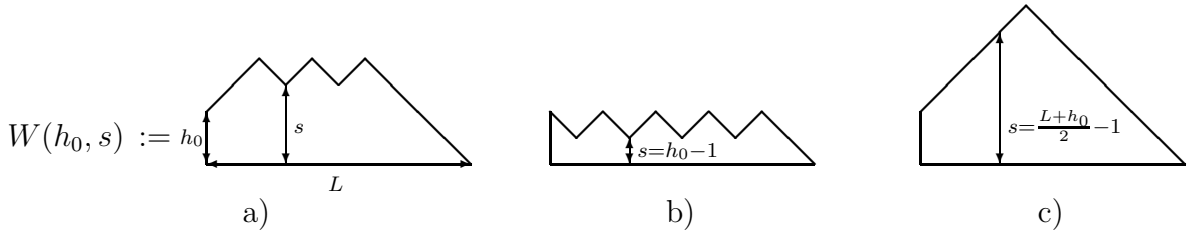
\begin{figure}
\hfil
$
\begin{array}{ccc}
W(h_0,s)\, :=
\begin{picture}(140,20)(-10,7)
\put(0,0){\vector(1,0){100}}
\put(100,0){\vector(-1,0){100}}
\put(46,-10){$\scriptstyle L$}
\put(0,0){\vector(0,1){20}}
\put(0,20){\vector(0,-1){20}}
\put(-10,8){$\scriptstyle h_0$}
\put(30,0){\vector(0,1){30}}
\put(30,30){\vector(0,-1){30}}
\put(33,12){$\scriptstyle s$}
{\thicklines
\put(0,20){\line(1,1){20}}\put(20,40){\line(1,-1){10}}
\put(30,30){\line(1,1){10}}\put(40,40){\line(1,-1){10}}
\put(50,30){\line(1,1){10}}\put(60,40){\line(1,-1){40}}
}
\end{picture}
&
\begin{picture}(130,20)(10,7)
\put(0,0){\line(1,0){100}}
\put(0,0){\line(0,1){20}}
\put(30,0){\vector(0,1){10}}
\put(30,10){\vector(0,-1){10}}
\put(33,3){$\scriptstyle s=h_0-1$}
{\thicklines
\put(0,20){\line(1,-1){10}}
\put(10,10){\line(1,1){10}}\put(20,20){\line(1,-1){10}}
\put(30,10){\line(1,1){10}}\put(40,20){\line(1,-1){10}}
\put(50,10){\line(1,1){10}}\put(60,20){\line(1,-1){10}}
\put(70,10){\line(1,1){10}}\put(80,20){\line(1,-1){20}}
}
\end{picture}
&
\begin{picture}(130,20)(10,7)
\put(0,0){\line(1,0){100}}
\put(0,0){\line(0,1){20}}
\put(30,0){\vector(0,1){50}}
\put(30,50){\vector(0,-1){50}}
\put(33,12){$\scriptstyle s={L+h_0\over 2}-1$}
{\thicklines
\put(0,20){\line(1,1){40}}\put(40,60){\line(1,-1){60}}
}
\end{picture}
\\[5mm]
\mbox{a)\hspace{6mm}}&\mbox{b)\hspace{18mm}}&\mbox{c)\hspace{18mm}}
\end{array}
$
\hfil
\caption{\footnotesize A typical configuration $W(h_0,s)$ is shown on picture a).
It is uniquely defined by its size $L$, its left boundary height $h_0$
and  the height $s$ of its contact points in the bulk.
A number of $s$-contacts in such configuration equals $({L+h_0\over 2}-s-1)$.
Boundary cases with $s=(h_0-1)$ and $s=({L+h_0\over 2}-1)$
are shown, respectively, on pictures b) and c).
In the last case one uses parameter $s$ just by analogy, because
configuration c) contains 0 number of contacts in the bulk.
}
\lb{generators}
\end{figure}

This Conjecture is graphically illustrated on Fig.6 on page
\pageref{FigAA}.
\medskip

Most of the orbits appearing in
the left hand sides of relations
(\ref{p-sums-a}) and (\ref{p-sums-b})
are singlets and doublets. Namely, one has singlet $(h_0-1)$-orbits
in case $s\geq h_0$ and singlet $h_0$-orbits for $s=h_0-1$;
one has doublet $h_0$-orbits in case $s\geq h_0+1$.
The larger size $2^{L-h_0\over 2}$-plet $h_0$- and $(h_0-1)$-orbits
appear, respectively,  for $s=h_0$ and $s=h_0-1$.

Relations for singlets and doublets provide explicit expressions for
a number of stationary state coefficients
in the model B. In particular,
formulas for $(h_0-1)$-singlets generated by elements
$W(h_0,s)$ with $s\geq h_0$
were observed  in \cite{MNGB}.

Noticing that two configurations from $h_0$-doublet
generated by $W(h_0,h_0+1)$
correspond to $(h_0-1)$- and $(h_0+2)$-singlets generated by
$W(h_0,h_0+1)$ and $W(h_0+2,h_0+1)$, respectively,
one concludes that expressions from the right hand sides of relations
(\ref{p-sums-a}) and (\ref{p-sums-b}) should comply certain consistency
rules. These are just Pascal's hexagon relations (\ref{hexagon1}),
(\ref{hexagon2}) and in this way Pascal's hexagon relations were
first time observed.
\medskip

We finish the section with several observations about detailed
normalizations and  extremes in the model B.

\vspace{3mm}\noindent {\bf Conjecture~9 (Model B: detailed distributions I).~}
{\em
For the quantities $S^{(b)}_{L,N}$ introduced in (\ref{MS-detailed})
one has formulas
\ba
\lb{Sb}
\begin{array}{rrcl}
\mbox{for $L$ even:}\hspace{5mm} &
S^{(b)}_{L,N} &=& S^{(a)}_{L+1, [{N+1\over 2}]}\;
S^{(a)}_{L, [{N\over 2}]}\, ,
\\[5mm]
\mbox{for $L$ odd:}\hspace{5mm} &
S^{(b)}_{L, N} &=& S^{(a)}_{L+1, [{N\over 2}]}\;
S^{(a)}_{L,[{N+1\over 2}]}\, .
\end{array}
\ea
}
The same formulas for $S^{(b)}_{L,N}$ were observed in \cite{MNGB}
(see eq.(48) there).

We remind that label $N$
refers to global minimum of Ballot paths  contributing to
distribution $S^{(b)}_{L,N}$ (it should be greater or equal to $N$).
$N$ may take all integer values from 0 to $L$.
For $N=0$  relations (\ref{Sb}) coincide with eq.(\ref{sigma2}).

\medskip
One can consider another set of detailed distributions
$\Sigma^{(b)}_{L,N,M}$ which
count stationary state contributions
of all configurations $w$ whose
left boundary height is fixed as $h_0(w)=N$ and
whose all $s$-contacts in the bulk satisfy condition
$s\geq M$ (i.e., whose global
minimum in the bulk is greater or equal to $M$).
For $M\geq N$
all such configurations can be obtained by adding tiles (not half-tiles!)
to the configuration $W(h_0=N,s=M)$ drawn on Fig.\ref{generators} a).

Note that
(in contrast to the case of $S^{(b)}_{L,N}$)
for distributions $\Sigma^{(b)}_{L,N,M}$
label $N$ may take only
integer values between zero and $L$ which are of the same parity as $L$.

\vspace{3mm}\noindent {\bf Conjecture~10
(Model B: detailed distributions II).~}
{\em
In case $M\geq N$  quantities $\Sigma^{(b)}_{L,N,M}$
are given by formula
\be
\lb{Sigma}
\Sigma^{(b)}_{L,N,M}\, =\, S^{(a)}_{L+1,\, M-[{N-1\over 2}]}\;
S^{(a)}_{L,[{N-1\over 2}]}\, .
\ee
}

We were unable to find  similar expressions in case $M<N$.
\medskip

The right hand sides of eqs.(\ref{Sb}) and (\ref{Sigma}) contain
the same square combination $S^{(a)}_{L+1,X} S^{(a)}_{L,Y}$ but
for indices $X, Y$ taking  values in adjacent domains. Hence,
formally speaking one can view distributions $S^{(b)}_{L,N-1}$ and
$S^{(b)}_{L,N-2}$ as extrapolations of distribution
$\Sigma^{(b)}_{L,N,M}$ to cases $M=N-1$ and $M=N-2$, respectively.
However, we don't know  any non-formal explanation to this fact.

\vspace{3mm}\noindent {\bf Conjecture~11 (Model B: detailed
extremes).~} {\em Among elements of the set $\{w\}_{Ballot}^N$
configuration with maximal possible number $[{L-N\over 2}]$ of
$N$-contacts enters the stationary state with the maximal
coefficient $M^{(b)}_{L,N}$ (see def.(\ref{MS-detailed})). All
such configurations are  generating elements $W(h_0,s=h_0-1)$, or
$W(h_0,s=h_0)$, respectively,  for  $h_0$-, or $(h_0-1)$-singlet
orbits. Hence, expressions for maxima $M^{(b)}_{L,N}$ are
particular cases of formulas (\ref{p-sums-a}),(\ref{p-sums-b})
(index $s$ there corresponds to $N$ in the notation for detailed
maxima). They are \ba \lb{chet} \mbox{for $L$ even:} &&
\begin{array}{rcl}
M^{(b)}_{L,N=2m}& =& S^{(a)}_{L,m-1}\; S^{(a)}_{L,m}\, ,
\\[2mm]
M^{(b)}_{L,N=2m+1}& =& S^{(a)}_{L+1,m+1}\; S^{(a)}_{L-1,m}\, ;
\end{array}
\\[5mm]
\lb{nechet}
\mbox{for $L$ odd:} &&
\begin{array}{rcl}
M^{(b)}_{L,N=2m}& =& S^{(a)}_{L+1,m}\; S^{(a)}_{L-1,m-1}\, ,
\\[2mm]
M^{(b)}_{L,N=2m+1}& =& S^{(a)}_{L,m}\; S^{(a)}_{L,m+1}\, .
\end{array}
\ea
In case of $L$ even and $N=0$ there is one more
configuration
$
w=\{h_0=2, h_i=\epsilon(i)\}
$
(see Fig.\ref{Fig1} b) on p.\pageref{Fig1})
which enters the stationary state with
the same maximal coefficient.

\medskip
Among all Ballot paths  with the same left boundary height
$h_0$
configuration which have no any contact points in the bulk
enters the stationary state with a minimal coefficient $m^{(b)}_{L,h_0}$.
All such configurations are  generating elements
$W(h_0,s={L+h_0\over 2}-1)$ of  $(h_0-1)$-singlet orbits.
So, expressions for $m^{(b)}_{L,h_0}$  can be found among formulas
eqs.(\ref{p-sums-a}), (\ref{p-sums-b}). They are
\be
\lb{det-min}
m^{(b)}_{L,h_0}\, =\, S^{(a)}_{L,\, [{h_0-1\over 2}]}\, .
\ee
}

\subsection{Model C: right orbits and factorizations to model B}

In this section we consider right 0- and 1-orbits in a family of
Anchored Cross paths and describe corresponding stationary
distributions for the model C. Our key
observations are following.
\medskip

The family of Anchored Cross paths splits uniquely into collection of all
(mutually nonintersecting) maximal right 0-/1-orbits.

The maximal right 0-orbits of length $L$
are generated by  elements
\be
\lb{ah3}
w=\{h_k\}_{k=0,\dots,L}:\quad
h_L  = 0\, .
\ee

The maximal right 1-orbits of length $L$
are generated by elements
\be
\lb{ah4}
w=\{h_k\}_{k=0,\dots,L}:\quad
\mbox{~either~} h_L =  0, \mbox{~or~} h_L=2 \mbox{~~and~}
\min_{0\leq i\leq L} h_i=1\; (\mbox{i.e.,~}\neq 0)\,  .
\ee

We further observe that sums of the stationary coefficients over maximal right
0- and 1-orbits in case C are related to certain
stationary coefficients in case B. These relations generalizing formula
(\ref{mu}) are given in

\vspace{2mm}\noindent{\bf Conjecture~12.}{\em
Take any Ballot paths $u$ and $v$ of the lengths $L$ and $(L+1)$,
respectively. Denote $v^*$ an Anchored Cross path of the length $L$
which is a (right) reduction of the Ballot path $v$, i.e.,
$$
\mbox{for all~~} k=0,\dots ,L,\quad h_k(v^*) =
\left\{
\begin{array}{rl}
h_k(v)+1\, , &\mbox{if~} \min_{k=0}^L h_k(v) = 0\, ,
\\[1mm]
h_k(v)-1\, , &\mbox{otherwise.}
\end{array}
\right.
$$
With these notations one has
\ba
\lb{77a}
\sum_{\nu\in{\cal O}_r(u,0)}p_L^{(c)}(\nu) &=&
\mbox{Denominator of~}\left({S^{(a)}_L\over S^{(a)}_{L+2}}\right)\,
p^{(b)}_L(u)\, \equiv\, {m^{(c)}_L S^{(a)}_{L+2}\over
S^{(a)}_L}\, p^{(b)}_L(u)\,,
\\[3mm]
\lb{77b} \sum_{\nu\in{\cal O}_r(v^*,1)}p_L^{(c)}(\nu) &=&
\mbox{Numerator of~}\left({S^{(a)}_L\over S^{(a)}_{L+2}}\right)\,
p^{(b)}_{L+1}(v)\, \equiv\, m^{(c)}_L\, p^{(b)}_{L+1}(v)\, . \ea
Here summation is taken over all elements of the right 0-, or
1-orbits generated by $u$, or $v^*$ (configurations $u$ and $v^*$
span all generating elements described in (\ref{ah3}),
(\ref{ah4})). }
\medskip

This Conjecture is graphically illustrated on Fig.7 on page
\pageref{FigB}. \bigskip

\noindent\hspace*{2mm}
\parbox{163mm}{
{\bf Remark 1.~}
{\footnotesize
In normalization (\ref{new-norm}) relations
(\ref{77a}) and (\ref{77b}) read
\ba
\lb{77c}
\sum_{\nu\in{\cal O}_r(u,0)}\tilde{p}_L^{(c)}(\nu) &=& S^{(a)}_{L+2}
p^{(b)}_L(u)\, ,
\\[1mm]
\lb{77d}
\sum_{\nu\in{\cal O}_r(v^*,1)}\tilde{p}_L^{(c)}(\nu) &=& S^{(a)}_L\,
p^{(b)}_{L+1}(v)\, .
\ea
In this presentation Conjecture 12 looks very much in the spirit of
Conjecture 7 and one can be immediately convinced that it
refines formula (\ref{mu2}).
}

\medskip
{\bf Remark 2.~}
{\footnotesize
In the model C  stationary state obeys  mirror symmetry. Namely,
any pair of configurations $\{h_k\}$ and $\{h'_k\}$ whose shapes are left-right
symmetric, i.e.,

\medskip
$h'_k = h_{L-k}$ for all $k$ in case of  $L$ even;

\medskip
$h'_k = \left\{\begin{array}{rcl}
h_{L-k} + 1\, ,& \mbox{if}& \min_{i=0}^L h_i =0\, ,
\\[1mm]
h_{L-k}-1\, ,& \mbox{if}& \min_{i=0}^L h_i =1\, ,
 \end{array}\right.$ for all $k$ in case of $L$ odd;

\medskip
enter the stationary state with the same coefficients.
Using this symmetry
one can reformulate Conjecture~12 in the language of left orbits.
}}
\medskip

There are many singlets
and doublets among the orbits appearing
in the right hand sides of
eqs.(\ref{77a}) and (\ref{77b}). E.g., all 0-orbits
generated by elements $w: h_L(w)=0$ and $h_k(w)\geq 1\;
\forall\; k=1,\dots ,L-1$, are doublets;
all 1-orbits generated by configurations $w:
h_L(w)=0$ are singlets. Relations for singlets and doublets provide
a number of explicit expressions for stationary state coefficients
of the model C.
In particular, expressions which are presented in a right column of Table 2
in \cite{MNGB} all correspond to the right 1-singlets.
To illustrate the practical use of Conjectures 12 and 8 we shall derive
here formulas for stationary state coefficients of configurations
\be
\lb{XX}
X(s)\, :=
\begin{picture}(210,45)(-27,7)
\put(0,0){\vector(1,0){140}}
\put(140,0){\vector(-1,0){140}}
\put(48,-12){$\scriptstyle L \mbox{\em ~--- even}$}
\put(0,0){\vector(0,1){20}}
\put(0,20){\vector(0,-1){20}}
\put(-23,7){$\scriptstyle h_0=2$}
\put(140,0){\vector(0,1){20}}
\put(140,20){\vector(0,-1){20}}
\put(145,7){$\scriptstyle h_L=2$}
\put(60,0){\vector(0,1){40}}
\put(60,40){\vector(0,-1){40}}
\put(65,18){$\scriptstyle s$}
{\thicklines
\put(0,20){\line(1,-1){10}}
\put(10,10){\line(1,1){40}}
\put(50,50){\line(1,-1){10}}
\put(60,40){\line(1,1){10}}
\put(70,50){\line(1,-1){10}}
\put(80,40){\line(1,1){10}}
\put(90,50){\line(1,-1){40}}
\put(130,10){\line(1,1){10}}
}
\end{picture}
\vspace{5mm}
\ee
where configuration's size
$L$ may take even values only and
parameter $s$ runs from 1 to $(L-2)\over 2$.
Denote
$$
\begin{array}{cc}
Y(s)\, :=
\begin{picture}(170,45)(-27,7)
\put(0,0){\vector(1,0){140}}
\put(140,0){\vector(-1,0){140}}
\put(48,-12){$\scriptstyle L \mbox{\em ~--- even}$}
\put(0,0){\vector(0,1){20}}
\put(0,20){\vector(0,-1){20}}
\put(-23,7){$\scriptstyle h_0=2$}
\put(60,0){\vector(0,1){40}}
\put(60,40){\vector(0,-1){40}}
\put(65,18){$\scriptstyle s$}
{\thicklines
\put(0,20){\line(1,-1){10}}
\put(10,10){\line(1,1){40}}
\put(50,50){\line(1,-1){10}}
\put(60,40){\line(1,1){10}}
\put(70,50){\line(1,-1){10}}
\put(80,40){\line(1,1){10}}
\put(90,50){\line(1,-1){50}}
}
\end{picture}
&
Z(s)\, :=
\begin{picture}(200,45)(-27,7)
\put(0,0){\vector(1,0){150}}
\put(150,0){\vector(-1,0){150}}
\put(52,-12){$\scriptstyle L \mbox{\em ~--- odd}$}
\put(0,0){\vector(0,1){30}}
\put(0,30){\vector(0,-1){30}}
\put(-23,7){$\scriptstyle h_0=3$}
\put(60,0){\vector(0,1){50}}
\put(60,50){\vector(0,-1){50}}
\put(65,18){$\scriptstyle s$}
{\thicklines
\put(0,30){\line(1,-1){10}}
\put(10,20){\line(1,1){40}}
\put(50,60){\line(1,-1){10}}
\put(60,50){\line(1,1){10}}
\put(70,60){\line(1,-1){10}}
\put(80,50){\line(1,1){10}}
\put(90,60){\line(1,-1){60}}
}
\end{picture}
\end{array}
\vspace{7mm}
$$
According to Conjectures 8 and 12 the
stationary state coefficients for these configurations and for configurations
$W(h_0=0,s)$ and $W(h_0=1,s)$ (see fig.\ref{generators} on
p.\pageref{generators})
satisfy relations
\ba
\nonumber
\mbox{\footnotesize right 0-doublet (eq.(\ref{77a})):}&&
p_L^{(c)}(Y(s)) + p_L^{(c)}(X(s))\, =\,
{m_L^{(c)} S^{(a)}_{L+2}\over S^{(a)}_L}\;
p_L^{(b)}(Y(s))\, ,\quad (s\geq 1)\, ,
\\[3mm]
\nonumber
\mbox{\footnotesize right 1-singlet (eq.(\ref{77b})):}&&
p_L^{(c)}(Y(s)) \, =\,
m_L^{(c)} \;
p_{L+1}^{(b)}(Z(s+1))\, ,
\\[3mm]
\nonumber
\mbox{\footnotesize left 0-doublet (first eq.(\ref{p-sums-a})):}&&
p_L^{(b)}(W(0,s)) + p_L^{(b)}(Y(s)) \, =\,
S^{(a)}_{L+1}\, S^{(a)}_{L-1,s}\, ,\quad (s\geq 1)\, ,
\\[3mm]
\nonumber
\mbox{\footnotesize left (-1)-singlet (second eq.(\ref{p-sums-a})):}&&
p_L^{(b)}(W(0,s))  \, =\,
S^{(a)}_{L}\, S^{(a)}_{L,s}\, ,
\\[3mm]
\nonumber
\mbox{\footnotesize left 1-doublet (second eq.(\ref{p-sums-b})):}&&
p_{L+1}^{(b)}(W(1,s+1)) + p_{L+1}^{(b)}(Z(s+1)) \, =\,
S^{(a)}_{L+2}\, S^{(a)}_{L,s}\, ,\quad (s\geq 1)\, ,
\\[3mm]
\nonumber
\mbox{\footnotesize left 0-singlet (first eq.(\ref{p-sums-b})):}&&
p_{L+1}^{(b)}(W(1,s+1))  \, =\,
S^{(a)}_{L+1}\, S^{(a)}_{L+1,s+1}\, ,
\ea
wherefrom one obtains
\be
\lb{X}
p_L^{(c)}(X(s))\, =\, m_L^{(c)}\left\{
{S^{(a)}_{L+2}\, S^{(a)}_{L+1}\, S^{(a)}_{L-1,s}\over S^{(a)}_L}\,
-\, 2\, S^{(a)}_{L+2}\, S^{(a)}_{L,s}\, +\, S^{(a)}_{L+1}\, S^{(a)}_{L+1,s+1}
\right\} .
\ee

In the last conjecture we describe  largest components of the
stationary states of the model C.

\vspace{2mm}\noindent{\bf Conjecture~13 (Model C: maxima).}{\em
The maximal coefficient $M_L^{(c)}$ appear in
the set $\{p_L^{(c)}\}$ with multiplicity 1 for $L$ even and 2 for
$L$ odd.

For $L$ odd one of the corresponding interface configurations
is the substrate
shown on Figure \ref{Fig1} c) on page \pageref{Fig1}. The second one
is a mirror image
of the first (see Remark 2 to Conjecture 12).
For $L$ even the corresponding configuration is $X(1)$
(see def.(\ref{XX})). Explicit values of the maximal coefficients
are
\ba
\lb{mc1}
\mbox{for $L$ odd:}&&
M^{(c)}_L\, =\,
m^{(c)}_L\, S^{(a)}_L\, S^{(a)}_{L+2,1}\, ,
\\[3mm]
\lb{mc2}
\mbox{for $L$ even:}&&
M^{(c)}_L\, =\, m_L^{(c)}\left\{
{S^{(a)}_{L+2}\, S^{(a)}_{L-1}\, S^{(a)}_{L+1,1}\over S^{(a)}_L}
\, -\, S^{(a)}_L\, S^{(a)}_{L+2,1}
\right\} .
\ea
Relation (\ref{mc2}) is a particular case of eq.(\ref{X})
simplified with the use of Pascal's hexagon relations.

A mirror image of the configuration $X(1)$, which is the
substrate, enters the stationary state with next to largest
coefficient \ba \mbox{for $L$ even:} &&
p_L^{(c)}(\mbox{Substrate})\, =\, m_L^{(c)}\, S^{(a)}_{L+1}\,
S^{(a)}_{L+1,1}\, . \ea }

\section{Discussion}

Two of the issues discussed in this paper deserve further investigation.

The first one is Pascal's hexagon relation which surprisingly
comes as defining recurrent equation for some of the RPM's
stationary coefficients. At the moment we do not have explanation
to this fact. A very preliminary investigation of Pascal's
hexagon relation is carried out in the Appendix. It shows
that besides the solution $S^{(a)}_{L,N}$ (\ref{Sa}) which is
related to RPM with open boundaries the relation admits also
solutions which manifest themselves in the periodic stochastic
models (see eq.(\ref{RQ}) and the sentence above it; see also
remark \ref{agaga} on page \pageref{agaga}).
Besides that, an extensive list of relations
(\ref{asm1})--(\ref{asm6})) obtained in the Appendix
indicates an existence of a profound relation
between solutions of Pascal's hexagon reccurence and
combinatorics of the alternating sign matrices.
This relation has to be further explored.

Secondly, while studying the stationary states of the models A, B and
C we observed close relations between their probability distributions.
Similar relations were also observed in Ref.\cite{GNiPR} for the spectra of
the models.
Such relations have been interpreted by introducing the notion of
orbits on the space of configurations of the models B and C.
These facts suggest an idea that models B
and A  can be obtained by factorization of the
model C. It would be interesting to find out such conjectural
factorization at the level of boundary extended Temperley-Lieb
algebras.

\section{Acknowledgements}
The author is grateful to  Jan de Gier, Andrey Mudrov,
Bernard Nienhuis, Yuri Stroganov and Nikolai Tyurin for useful discussions.
Special thanks to Vladimir Rittenberg whose constant interest in the subject
and valuable advices provide a lot of inspiration to this work.
The work is supported in part by RFBR grant \# 03-01-00781 and by
the grant of Heisenberg-Landau Foundation.
The author gratefully acknowledges warm hospitality of
Physikalisches Institut of the University of Bonn
and of  Lorentz Center at the University of Leiden
where this work was completed.

\vspace{9mm}
\renewcommand{\appendixname}{{\noindent\Large\bf Appendix.~ Solutions of
Pascal's hexagon relation and\\[1mm]
\phantom{a}\hspace{30mm}
ASM numbers\vspace{5mm}}}
\appendix
\appendixname
\renewcommand{\theequation}{{\small A}.\arabic{equation}}
\setcounter{equation}0

Consider a set of numbers  which are placed on
vertices of a planar trigonal lattice.
Introducing coordinate lines on the lattice
as it is shown on figure (\ref{coord})
we label the numbers in the set by a
pair of integers: $\{f_{m,n}\}$, $m,n\in {\Bbb Z}$.
\be
\lb{coord}
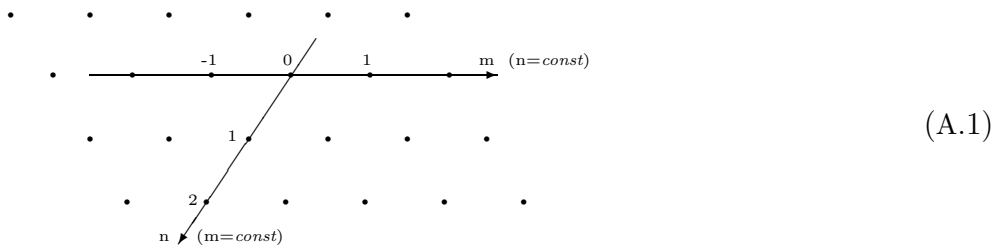
\begin{array}{c}
\begin{picture}(50,90)
\put(-82,62){\vector(1,0){154}}
\put(3.5,76){\vector(-2,-3){52}}
\multiput(-68,14)(30,0){6}{\circle*{2}}
\multiput(-82,38)(30,0){6}{\circle*{2}}
\multiput(-96,62)(30,0){6}{\circle*{2}}
\multiput(-112,85)(30,0){6}{\circle*{2}}
\put(-40,66){\tiny -1}
\put(-9,66){\tiny 0}
\put(21,66){\tiny 1}
\put(65,66){\tiny m~~(n={\tiny\em const})}
\put(-30,37){\tiny 1}
\put(-45,13){\tiny 2}
\put(-56,-1){\tiny n~~~~(m={\tiny\em const})}
\end{picture}
\end{array}
\ee
For any six  numbers $a, a', b, b', c, c'$  in the set
which are situated  on apices of a hexagon
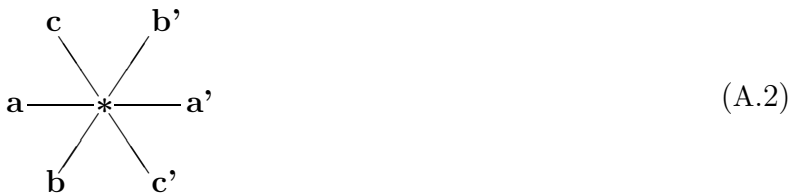
\be
\lb{hexagon}
\begin{array}{c}
\begin{picture}(50,60)(0,-25)
\put(31,-2.5){\bf a'}
\put(-37,-2.5){\bf a}
\put(18,28){\bf b'}
\put(-22,28){\bf c}
\put(-22,-33){\bf b}
\put(18,-33){\bf c'}
\put(-3,-6.5){\bf *}
\put(4,0){\line(1,0){25}}
\put(-4,0){\line(-1,0){25}}
\put(2,3.4){\line(2,3){15}}
\put(2,-3.4){\line(2,-3){15}}
\put(-2,3.4){\line(-2,3){15}}
\put(-2,-3.4){\line(-2,-3){15}}
\end{picture}
\\
\end{array}
\ee
we impose a condition
\be
\lb{pascal}
a a' + b b' = c c'\, ,
\ee
which can be viewed as a generalization of Pascal's triangle relation to
the case of hexagon. In coordinatization (\ref{coord}) it reads
\be
\lb{pascal2}
f_{m-1,n}\, f_{m+1,n}\, +\, f_{m,n-1}\, f_{m,n+1}\, =\,
f_{m-1,n-1}\, f_{m+1,n+1}\, .
\ee

As the author has learned from Yuri Stroganov Pascal's hexagon
relations belong to a wide family of bilinear equations, including
Somos sequences,  number walls, S-arrays,
cube reccurence  etc., which are currently extensively investigated
in algebraic combinatorics (see, e.g.,
\cite{FZ} and references therein; see also bilinear forum on
the web). Moreover, it has been  mentioned in physical literature
under the name of discrete Boussinesq equation (see
eq.(8.11) in \cite{Z} and the references therein).
In \cite{Z} it was obtained  as certain  2-dimensional reduction of the Hirota's
difference equation, which is also called the
octahedron reccurence by combinatorialists.

A particular solution of Pascal's hexagon relation is given by formulas
\ba
\lb{Saa}
\hspace*{-7mm}
f_{m,n}\, =\,
2^{^{-[n^2/ 4]}}
\prod_{p=1}^{n}{1\over (2p-1)!!}\,
\prod_{p=0}^{[{n-1\over 3}]}{\bigl(m-[{n+p\over 2}]-p-1\bigr)!\over
\bigl(m-2n+3p\bigr)!}\,
\prod_{p=0}^{[{n-2\over 3}]}{\bigl(2m+2n-6p-3\bigr)!!\over
\bigl(2m-2[{n+p\over 2}]+4p+1\bigr)!! }\, ,
\\[5mm]
\hspace*{-7mm}
\lb{Sab}
f_{-m,-n}\, =\,
2^{^{-[n^2/4]}}
\prod_{p=1}^{n-1}{1\over (2p-1)!!}\,
\prod_{p=0}^{[{n\over 3}]-1}{\bigl(m-[{n+p\over 2}]-p-2\bigr)!\over
\bigl(m-2n+3p+2\bigr)!}\,
\prod_{p=0}^{[{n-2\over 3}]}{\bigl(2m+2n-6p-5\bigr)!!\over
\bigl(2m-2[{n+p\over 2}]+4p-1\bigr)!! }\, ,
\ea
where $m$ is an integer  and  $n$ is a  nonnegative integer.
Here for negative values of $m$  ratios of factorials are  treated as
Pochhammer symbols:
$$
(n-1)!/(k-1)! =\prod_{j=k}^{n-1} j\, =: (k)_{(n-k)}\, ,\qquad
(2n-1)!!/(2k-1)!! = 2^{(n-k)}(k)_{(n-k)}(2k)_{(2n-2k)}\, .
$$
The corresponding distribution of numbers on the lattice
is shown in table \ref{tab-S} on page \pageref{tab-S}.

\medskip
Formulas  (\ref{Saa}) and (\ref{Sab})
can be reproduced in a following way.
First, one  assigns certain initial data to $f_{m,n}$
(like in Pascal's triangle case)
\be
\lb{boundf}
f_{m\geq o,n=0}=1\, , \qquad
f_{m\geq 0, n=-1}=1\, , \qquad
f_{m=2n-1, n> 0}=0\, ,
\ee
and  calculates values of $f_{m,n}$ for $n> 0$ and $m\geq 2n$
row by row (i.e., consecutively for $n=1,2,3,$ etc.)
using relations (\ref{pascal2}). In this way formula (\ref{Saa})
for detailed distributions $S^{(a)}_{L,n}=f_{m=L,n}$ was
obtained in Sec. 3.3.

Then, one observes that formula (\ref{Saa}) is well defined
on a half-plane  $n\geq 0$ and
gives numbers satisfying Pascal's hexagon relation
in this sector.
Note that combinatorial functions $R(n,L)$ and $Q(n,L)$
introduced in \cite{MNGB} (see eqs.(16) and (17) there)
both are related to function (\ref{Saa})
\be
\lb{RQ}
f_{m,n} = R(n+1,m+1)\, , \qquad
f_{-m,n} =(-1)^{[{n+1\over 2}]} Q(n+1,m+n+1)\, .
\ee
Note also that
in the sector
$m\leq 2n-1$,~ $n\leq 2m-1$
numbers $f_{m,n}$ vanish
(see tab.(\ref{tab-S})) that, in particular, destroys the connection
between numbers $f_{m,n}$ in sectors  $m\geq 2n$,~ $n\geq 0$
and  $2m\leq n$,~ $n\geq 0$ via Pascal's hexagon relations.

Further on,  comparing  numbers $f_{m,n}$ in
sectors $m\geq 2n\geq 0$ and $n\geq 2m\geq 0$
one observes $m\leftrightarrow n$ symmetry
\be
\lb{mn-symm}
f_{m,n} = (-1)^{(mn\, +\, [4(m+n)/ 3])}\, 2^{-[(m-2n)/ 3]}\, f_{n,m}\,
\quad \forall\; m\geq n\geq 0\, .
\ee
Extrapolating this symmetry relation one  defines numbers
$f_{m,n}$ in sector $m\geq 0\geq n$
through those from sector $n\geq 0\geq m$. The Pascal's hexagon relation for the newly
defined numbers stays valid.

Finally, one can guess general formula
(\ref{Sab}) for $f_{m,n}$ on a half-plane $n<0$
analyzing the numbers in  sector $m\geq 0\geq n$. The
numbers $f_{m,n}$ thus obtained
satisfy relation (\ref{pascal2})
on the whole lattice.
Let us end up discussion of  the
solution (\ref{Saa}), (\ref{Sab})
with a list of remarks.
\begin{enumerate}
\item
$f_{m,n}$ take particularly simple values along the following directions
(see tab.\ref{tab-S})
\ba
\lb{simple1}
&&
\hspace{-20mm}
f_{m,0}\, =\, f_{m,-1}\, =\, 1
\\[2mm]
\lb{simple2}
&&
\hspace{-20mm}
f_{0,n}\, =\, (-1)^{[4n/3]}\, 2^{[n/3]}\, , \quad
f_{0,-n}\, =\, (-1/2)^{[2n/3]}\, , \quad \forall\; n\geq 0\, ,
\\[2mm]
\lb{simple3}
&&
\hspace{-20mm}
f_{-1,n}\, =\, (-2)^{[(n+2)/3]}\, , \quad
f_{-1,-n}\, =\, (-1)^{[(2-n)/3]}\, (1/2)^{[(2n-1)/3]}\, , \quad \forall\; n\geq 1\, ,
\\[2mm]
\lb{simple4}
&&
\hspace{-20mm}
f_{m=2n-1,n>0}\, =\, f_{m>0,n=2m-1}\, =\,
f_{m=2n+3.n<-2}\, =\, f_{m<-2,n=2m+3}\, =\, 0\, .
\ea
Conditions (\ref{simple1})--(\ref{simple4}) can be used as initial data
for reconstruction of all the nonzero numbers  $f_{m.n}$
by means of relation (\ref{pascal2}).
\item
Symmetry relations (\ref{mn-symm}) are valid for any pair $m$, $n$ such that $m\geq n$
(but not for $m<n$).
\item
The numbers $f_{m,n}$ given by eq.(\ref{Saa}) are always integer.
The numbers $f_{-m,-n}$ given by eq.(\ref{Sab}) become integer upon multiplication
by $2^{(n-1)}$. This fact is a particular manifestation of the
Laurent property, which was proved in general in \cite{FZ} (for specialization to
the case of octahedron recurrence see  \cite{S}).
\footnote{The author is grateful to David Speyer for pointing out
this fact to his attention.}
\item
\lb{agaga} The numbers $f_{m,n}$ in sectors $m\geq 2n\geq 0$ and
$0\leq n\leq 1-m$ are known to be related to open and periodic one
dimensional stochastic models, respectively (see \cite{MNGB} and
Sec.3.3). One may expect that there exists some one dimensional
stochastic process related to the numbers $f_{m,n}$ in sector
$n\leq -1$, $m\leq 2n+2$.
\end{enumerate}

Next, we consider solutions of Pascal's
hexagon relation  which are polynomials in one, or several variables.
Evaluated at particular values of their variable(s) they produce
numeric solutions to (\ref{pascal2}).
We found two examples of polynomial solutions.

First one is a set of polynomials in two variables $F_{m,n}(x,y)$
which are defined in the  sector $n\geq 0$, $m\geq 2n-1$.
They satisfy Pascal's hexagon relation (\ref{pascal2})  (just substitute $f_{m,n}$
by $F_{m,n}(x,y)$ there) and one can uniquely define them provided
that polynomials on the boundaries of the sector are fixed. We choose
\be
\lb{boundF}
F_{m>0,0}=1\, , \qquad
F_{m>1,1}=y+(m-2)x\, ,\qquad
F_{m=2n-1,n>0}=0\, ,
\ee
which is a generalization of boundary conditions (\ref{boundf}).
A nontrivial fact is that with such choice of the boundary conditions one
obtains polynomial (rather then rational) expressions for $F_{m,n}(x,y)$.
In  table \ref{tab-F} on page \pageref{tab-F} some
polynomials $F_{m,n}(x,y)$ for small values of $m$ and $n$
are listed.

The functions $F_{m,n}(x,y)$ obey rescaling symmetries\footnote{
Rescaling properties (\ref{rescale1}), (\ref{rescale2}) are
particular manifestations of ``gauge'' transformations on the set
of solutions of Pascal's hexagon equation. Namely, once a solution
$\{a,a',b,b',c,c'\}$ of  eq.(\ref{pascal}) is given a family of
gauge equivalent solutions $\{\alpha a,\alpha' a',\beta b,\beta'
b', \gamma c,\gamma' c'\}$ is parametrized  by six scaling numbers
$\alpha$, $\alpha'$, $\beta$, $\beta'$, $\gamma$ and $\gamma'$
satisfying  constraints $\alpha\alpha' = \beta\beta' =
\gamma\gamma' \neq 0$. On a trigonal lattice  this gauge
transformation is uniquely defined provided one fixes 6 mutually
independent scaling parameters for the vertices situated as shown
on a figure \ba \nonumber
\begin{array}{c}
\begin{picture}(50,28)
\multiput(2,0)(30,0){3}{\circle*{2}}
\multiput(-12,27)(30,0){3}{\circle*{2}}
\end{picture}
\end{array}
\ea
}
\ba
\lb{rescale1}
F_{m,n}(x,y) &=&
y^n\, F_{m,n}({x/y},1)\, , \qquad \mbox{for}\; y\neq 0\, ,
\\[2mm]
\lb{rescale2}
F_{m,n}(x,0) &=&
x^n\, F_{m,n}(1,0)\, ,
\\[2mm]
\lb{rescale3} F_{2n,n}(x,y) &=& y F_{2n,n-1}(x,y)\, . \ea It
follows then that only two values  $y=0$ and $y=1$ are essential.
We keep using variable $y$ as it is suitable for producing
integral solutions to Pascal's hexagon relations. In
particular, one finds following relations between polynomials
$F_{m,n}$ and the numbers given by formulas (\ref{Saa}) and
(\ref{Sab}) \ba F_{m+1,n}(1,0)\, =\, F_{m+2,n+1}(0,1) \, =\,
F_{m,n}(1,1) &=& f_{m,n}\, ,
\\[2mm]
F_{m,n}(2,3)&=& 2^n f_{-m,-n-1}\, .
\ea

\medskip
Another set of polynomials $G_{m,n}(x)$
are solutions of a variant
of Pascal's hexagon relations
\be
\lb{pascal3}
G_{m,n+1}G_{m,n-1} + G_{m+1,n-1} G_{m-1,n+1} \, =\, G_{m-1,n} G_{m+1,n}\, .
\ee
These relations result from a   different coordinatization of
Pascal's hexagon rules (\ref{hexagon}), (\ref{pascal}).
The coordinate frame for them
looks as
\be
\lb{coord2}
\begin{array}{c}
\begin{picture}(50,90)
\put(-82,62){\vector(1,0){154}}
\put(-18,76){\vector(2,-3){52}}
\multiput(-67,14)(30,0){6}{\circle*{2}}
\multiput(-82.5,38)(30,0){6}{\circle*{2}}
\multiput(-98,62)(30,0){6}{\circle*{2}}
\multiput(-112,85)(30,0){6}{\circle*{2}}
\put(-40,66){\tiny -1}
\put(-9,66){\tiny 0}
\put(21,66){\tiny 1}
\put(65,66){\tiny n~~(m={\tiny\em const})}
\put(0,37){\tiny 1}
\put(15,13){\tiny 2}
\put(40,-1){\tiny m~(n={\tiny\em const})}
\end{picture}
\end{array}
\ee
For this set of functions
we impose  boundary conditions
\be
\lb{boundG}
G_{0,n> 0}\, =\, 1\, , \quad
G_{1,n\geq 0}\, =\, 1+nx\, , \quad
G_{m>0,0}\, =\, x^{[(m+1)/3]} (x+1)^{[m/3]} (-1)^{[m/3]+[m/2]}\, ,
\ee
which allow one to iteratively reconstruct polynomials $G_{m,n}(x)$
(again, a nontrivial fact)
in the sector $m\geq 0$, $n\geq 0$.
In particular, one finds
\be
\lb{boundG2}
G_{m,1}\, =\, x^{[m/3]}\, (x+1)^{[(m+2)/3]}\, (-1)^{[(m+2)/3]+[(m+1)/2]}\, .
\ee
Boundary conditions (\ref{boundG}), (\ref{boundG2}) are generalizations
of the boundary data for $f_{m,n}$ along directions
$m<0,\ n\in\{0,1\}$,~  and $n>0,\ m\in\{0,-1\}$
(see table \ref{tab-S} and formulae (\ref{simple1})--(\ref{simple3})).
Explicit expressions for $G_{m,n}(a)$ for small values of $m$ and $n$ are given
in table \ref{tab-F} on page \pageref{tab-F}.

The functions $G_{m,n}(x)$ obey reflection symmetry
\ba
\lb{reflect}
G_{m,n}(x) & =& x^{[(m-n+1)/3]}\, {\tilde x}^{-[({\tilde m}-{\tilde n}+1)/3]]}\,
(-1)^{[(m+n)/2]}\, G_{{\tilde m},{\tilde n}}({\tilde x})\, , \quad \mbox{for}\;
x\neq 0,-1,
\\[2mm]
\lb{reflect2}
G_{m,n}(0) &=& (-1)^{m+1}\, G_{m-1,n+1}(-1)\, ,
\ea
where ${\tilde x} = -1-x$,~ ${\tilde m}=n-1$~ and~ ${\tilde n}=m+1$.
At certain values of $x$ they reproduce numbers (\ref{Saa})
\be
\lb{Gf}
G_{m,n}(0)\,  =\, f_{m+n-1,m-1}\, , \quad
G_{m,n}(1)\, =\, (-1)^{[(m+1)/2]}\, f_{-n,m}\, .
\ee

\medskip
Finally, it is remarkable that the solutions of Pascal's
hexagon relation reproduce in several ways  numbers which were discovered in
counting
various symmetry classes of the alternating
sign matrices (see \cite{R,K,B}). Below we present list of such formulae.
Relations (\ref{asm1}), (\ref{asm2}) and (\ref{asm3})--(\ref{asm4})
were observed  in \cite{MNGB} (see eqs.(18)--(23) there).
First equalities in (\ref{asm1}) and (\ref{asm5}) are particular
manifestations of eq.(\ref{rescale3}).
\ba
\nonumber
F_{2n,n-1}(1,1) &=& F_{2n,n}(1,1)\, =\,  G_{n+1,n}(0)\, =\, G_{n,n+1}(0)
\\[3mm]
\lb{asm1}
&&\hspace{25mm}=\,
\{1,3,26,646,\dots\}\, =\, A^V_{2n+1}\, ,\qquad n>0,
\\[3mm]
\lb{asm2}
F_{2n-1,n-1}(1,1) &=&
G_{n,n}(0)\, =\,
\{1,1,2,11,170,\dots\}\, =\, {A^{VH}_{4n\pm 1}\over A^V_{2n\pm 1}}\, ,
\qquad\;\;\;\;\, n\geq 0,
\\[3mm]
\lb{asm5}
F_{2n,n-1}(2,3) &=& {1\over 3} F_{2n,n}(2,3)\, =\,
\{1,7,143,8398,\dots\}\, =\, {A_{2n-1}\over A^V_{2n-1}}\, ,\quad\; n>0,
\\[3mm]
F_{2n+1,n}(2,3) &=& \{1,5,66,2431,252586,\dots\}\, =\, {A^{HT}_{4n}\over A_{2n}}\,
{A^V_{2n\pm 1}\over A^{VH}_{4n\pm 1}}\, ,\qquad\;\;\, n\geq 0,
\\[3mm]
\lb{asm3}
G_{n,n}(1) &=&
\{1,2,10,140,5544,\dots\}\, =\, A^{HT}_{2n}\, ,\qquad\qquad\quad n\geq 0,
\\[3mm]
G_{n,n+1}(1) &=&
\{1,3,25,588,39204,\dots\}\, =\, A^{HT}_{2n+1}\, ,\qquad\qquad n\geq 0,
\\[3mm]
G_{n+1,n}(1) &=&
\{1,2,2\!\times\! 7,7\!\times\! 42,42\!\times\! 429,\dots\}\,
=\,A_n A_{n+1}\, ,\;\;\; n\geq 0,
\\[3mm]
\lb{asm4}
G_{n,n+2}(1) &=&
\{1,2^2,7^2,42^2,429^2,\dots\}\, =\, (A_{n+1})^2\, ,\qquad\quad\;\; n\geq 0,
\\[3mm]
G_{n,n}(-{1\over 2}) &=&
\{1,{1\over 2},{1\over 2^2},{2^2\over 2^3},{6^2\over 2^4},
{33^2\over 2^5},{286^2\over 2^6}\dots\}\, =\,
{(A^{VH}_{2n+1})^2\over 2^n}\, ,\qquad\quad n\geq 0,
\\[3mm]
G_{n+1,n}(-{1\over 2}) &=&
\{1,{1\over 2},{2\over 2^2},{2\!\times\! 6\over 2^3},{6\!\times\! 33\over 2^4},
{33\!\times\! 286\over 2^5}\dots\} =
{A^{VH}_{2n+1}A^{VH}_{2n+3}\over 2^n}\, ,\;\; n\geq 0,
\\[3mm]
G_{n,n+1}(-{1\over 2}) &=&
\{1,0,{1\over 2^2},0,{3^4\over 2^4},0,{26^4\over 2^6},\dots\}\, =\,
\left\{\begin{array}{ll}
0 & \mbox{for $n\geq 1$ odd,}
\\
{(A^V_{2n+1})^4\over 2^n} & \mbox{for $n\geq 0$ even,}
\end{array}\right.
\\[3mm]
\nonumber
G_{n+2,n}(-{1\over 2}) &=& \{{1\over 2},{1^3\over
2^2},{1^3\!\times\! 3\over 2^3},{1\!\times\! 3^3\over 2^4},
{3^3\!\times\! 26\over 2^5}, {3\!\times\! 26^3\over 2^6}, \dots\}
\\[3mm]
\lb{asm6}
 &&\hspace{12mm} =\, {1\over
2^{n+1}}(A^V_{4[(n+2)/4]+1})^{p(n+2)}(A^V_{4[n/4]+3})^{p(n)}\, ,
\;\quad n\geq 0, \ea where  $p(n)=1,3,3,1,$~ if~ $(n-4[{n\over
4}])=0,1,2,3,$~ respectively.

Here $A_n$ is a total number of $n\times n$ alternating sign  matrices.
Notations  $A^V_n$, $A^{VH}_n$ and $A^{HT}_n$
stand, respectively,
for  numbers of vertically symmetric, vertically and horizontally symmetric
and half turn symmetric $n\times n$ alternating sign matrices.
Expressions for $A^V_{2n+1}$ and $A^{VH}_{4n\pm 1}/A^V_{2n\pm 1}$
are given in (\ref{sigma1}) and (\ref{sigma3}). Formulas for $A_n$
and $A^{HT}_n$ are
\be
A_n = \prod_{k=0}^{n-1}{(3k+1)!\over (n+k)!}\, ,\quad\;
A^{HT}_{2n} = (A_n)^2\prod_{k=0}^{n-1} {3k+2\over 3k+1}\, ,\quad\;
A^{HT}_{2n+1} = \prod_{k=1}^n {4\over 3} \left({(3k)!k!\over (2k)!^2}\right)^2\, .
\ee

\begin{figure}
$
\begin{array}{ll}
\begin{picture}(235,265)(-100,0)
\put(25,265){\underline{$0$-orbit, octet, $\mbox{sum over orbit} =
170 \times 11$}}

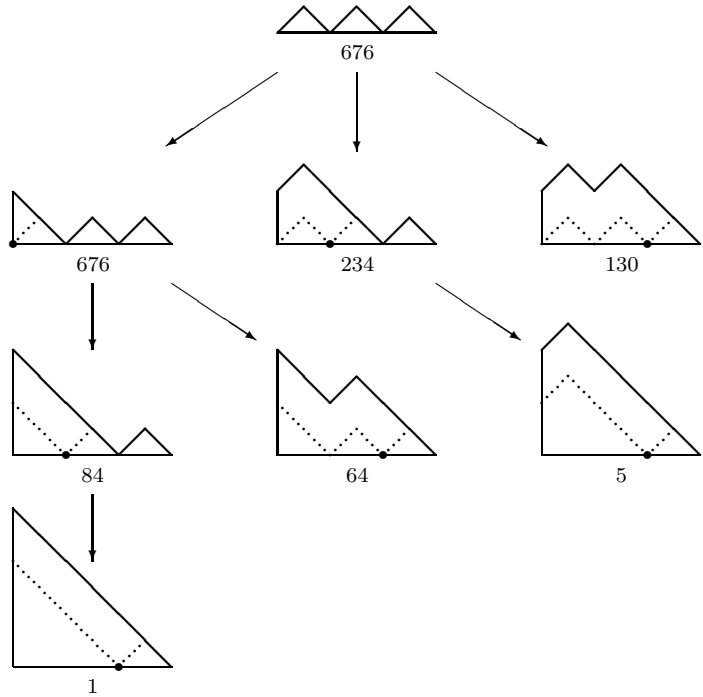
\put(100,240){\line(1,0){60}} {\thicklines
\put(100,240){\line(1,1){10}}\put(110,250){\line(1,-1){10}}
\put(120,240){\line(1,1){10}}\put(130,250){\line(1,-1){10}}
\put(140,240){\line(1,1){10}}\put(150,250){\line(1,-1){10}} }
\put(100,225){\vector(-3,-2){42}} \put(160,225){\vector(3,-2){42}}
\put(130,225){\vector(0,-1){30}} \put(124,230){\scriptsize 676}

\put(0,160){\line(1,0){60}} \put(0,160){\line(0,1){20}}
{\thicklines
\put(0,180){\line(1,-1){20}}\put(20,160){\line(1,1){10}}
\put(30,170){\line(1,-1){10}}\put(40,160){\line(1,1){10}}
\put(50,170){\line(1,-1){10}} }
\multiput(0,160)(2,2){5}{\circle*{.5}} \put(0,160){\circle*{3}}
\put(60,145){\vector(3,-2){32}} \put(30,145){\vector(0,-1){25}}
\put(24,150){\scriptsize 676}

\put(100,160){\line(1,0){60}} \put(100,160){\line(0,1){20}}
{\thicklines
\put(100,180){\line(1,1){10}}\put(110,190){\line(1,-1){30}}
\put(140,160){\line(1,1){10}}\put(150,170){\line(1,-1){10}}  }
\multiput(100,160)(2,2){5}{\circle*{.5}}
\multiput(110,170)(2,-2){5}{\circle*{.5}}
\multiput(120,160)(2,2){5}{\circle*{.5}}
\put(120,160){\circle*{3}} \put(160,145){\vector(3,-2){32}}
\put(124,150){\scriptsize 234}

\put(200,160){\line(1,0){60}} \put(200,160){\line(0,1){20}}
{\thicklines
\put(200,180){\line(1,1){10}}\put(210,190){\line(1,-1){10}}
\put(220,180){\line(1,1){10}}\put(230,190){\line(1,-1){30}}  }
\multiput(200,160)(2,2){5}{\circle*{.5}}
\multiput(210,170)(2,-2){5}{\circle*{.5}}
\multiput(220,160)(2,2){5}{\circle*{.5}}
\multiput(230,170)(2,-2){5}{\circle*{.5}}
\multiput(240,160)(2,2){5}{\circle*{.5}}
\put(240,160){\circle*{3}} \put(224,150){\scriptsize 130}

\put(0,80){\line(1,0){60}} \put(0,80){\line(0,1){40}}{\thicklines
\put(0,120){\line(1,-1){40}}\put(40,80){\line(1,1){10}}
\put(50,90){\line(1,-1){10}} }
\multiput(0,100)(2,-2){10}{\circle*{.5}}
\multiput(20,80)(2,2){5}{\circle*{.5}} \put(20,80){\circle*{3}}
\put(30,65){\vector(0,-1){25}} \put(26,70){\scriptsize 84}

\put(100,80){\line(1,0){60}}
\put(100,80){\line(0,1){40}}{\thicklines
\put(100,120){\line(1,-1){20}}\put(120,100){\line(1,1){10}}
\put(130,110){\line(1,-1){30}} }
\multiput(100,100)(2,-2){10}{\circle*{.5}}
\multiput(120,80)(2,2){5}{\circle*{.5}}
\multiput(130,90)(2,-2){5}{\circle*{.5}}
\multiput(140,80)(2,2){5}{\circle*{.5}} \put(140,80){\circle*{3}}
\put(126,70){\scriptsize 64}

\put(200,80){\line(1,0){60}}
\put(200,80){\line(0,1){40}}{\thicklines
\put(200,120){\line(1,1){10}}\put(210,130){\line(1,-1){50}} }
\multiput(200,100)(2,2){5}{\circle*{.5}}
\multiput(210,110)(2,-2){15}{\circle*{.5}}
\multiput(240,80)(2,2){5}{\circle*{.5}} \put(240,80){\circle*{3}}
\put(228,70){\scriptsize 5}

\put(0,0){\line(1,0){60}} \put(0,0){\line(0,1){60}} {\thicklines
\put(0,60){\line(1,-1){60}} }
\multiput(0,40)(2,-2){20}{\circle*{.5}}
\multiput(40,0)(2,2){5}{\circle*{.5}} \put(40,0){\circle*{3}}
\put(28,-10){\scriptsize 1}
\end{picture}
&
\\[5mm]
\begin{picture}(80,220)(0,0)
\put(2,205){\underline{$2$-orbit,  quartet, $ \mbox{sum over
orbit}= 50\times 4$}} \put(0,160){\line(1,0){60}}
\put(0,160){\line(0,1){20}} {\thicklines
\put(0,180){\line(1,1){10}}\put(10,190){\line(1,-1){10}}
\put(20,180){\line(1,1){10}}\put(30,190){\line(1,-1){30}}  }
\put(60,145){\vector(3,-2){32}} \put(30,145){\vector(0,-1){25}}
\put(24,150){\scriptsize 130}

\put(0,80){\line(1,0){60}} \put(0,80){\line(0,1){40}}{\thicklines
\put(0,120){\line(1,-1){20}}\put(20,100){\line(1,1){10}}
\put(30,110){\line(1,-1){30}} }
\multiput(0,100)(2,2){5}{\circle*{.5}} \put(0,80){\circle*{3}}
\put(30,65){\vector(0,-1){25}} \put(26,70){\scriptsize 64}

\put(100,80){\line(1,0){60}}
\put(100,80){\line(0,1){40}}{\thicklines
\put(100,120){\line(1,1){10}}\put(110,130){\line(1,-1){50}} }
\multiput(100,100)(2,2){5}{\circle*{.5}}
\multiput(110,110)(2,-2){5}{\circle*{.5}}
\multiput(120,100)(2,2){5}{\circle*{.5}} \put(120,80){\circle*{3}}
\put(128,70){\scriptsize 5}

\put(0,0){\line(1,0){60}} \put(0,0){\line(0,1){60}} {\thicklines
\put(0,60){\line(1,-1){60}} }
\multiput(0,40)(2,-2){10}{\circle*{.5}}
\multiput(20,20)(2,2){5}{\circle*{.5}} \put(20,0){\circle*{3}}
\put(28,-10){\scriptsize 1}
\end{picture}
&
\begin{picture}(80,220)(0,0)
\put(2,205){\underline{$4$-orbit,  doublet, $ \mbox{sum over
orbit}= 6\times 1$}}

\put(0,120){\line(1,0){60}}
\put(0,120){\line(0,1){40}}{\thicklines
\put(0,160){\line(1,1){10}}\put(10,170){\line(1,-1){50}} }
\put(30,105){\vector(0,-1){25}} \put(28,110){\scriptsize 5}

\put(0,40){\line(1,0){60}} \put(0,40){\line(0,1){60}} {\thicklines
\put(0,100){\line(1,-1){60}} }
\multiput(0,80)(2,2){5}{\circle*{.5}} \put(0,40){\circle*{3}}
\put(28,30){\scriptsize 1}
\end{picture}
\\[5mm]
\end{array}
$

\lb{FigAA} \caption{\bf Examples of maximal left orbits for size
$L=6_{_{_{}}}$ Ballot paths. \hspace{23mm}\rm Arrows on pictures
point from source configurations to their coverings; bold dots
indicate points where coverings happen; dashed lines show
configuration before covering happen. Numbers standing below
configurations are the corresponding stationary state coefficients
of the model B. Sums of the coefficients over orbits are
described by the first one of relations (\ref{p-sums-a}). }
\end{figure}

\begin{figure}
$
\begin{array}{c}
\begin{picture}(235,270)(-90,0)
\put(-78,270){\underline{$0$-orbit, octet, $\mbox{sum over orbit}
=\left(m_5^{(c)} S^{(a)}_7/ S^{(a)}_5\right)\times
p_5^{(b)}\left(\rule{0pt}{4mm}\right.$\phantom{\hspace{42mm}}}}
\put(206,268){\line(1,0){50}}\put(206,268){\line(0,1){10}}
\put(206,278){\line(1,-1){10}}\put(216,268){\line(1,1){10}}
\put(226,278){\line(1,-1){10}}\put(236,268){\line(1,1){10}}
\put(246,278){\line(1,-1){10}}
\put(258,270){$\left.\rule{0pt}{4mm} \right) = 170 \times 78$}

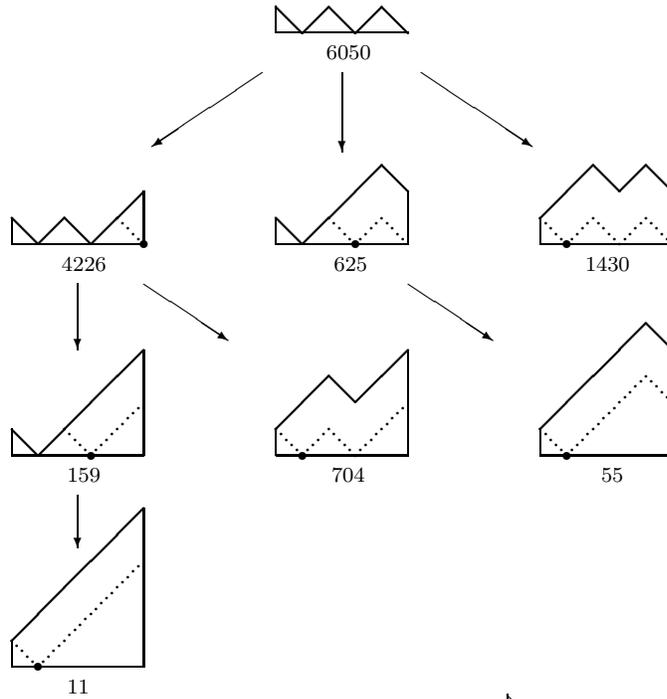
\put(100,240){\line(1,0){50}}
\put(100,240){\line(0,1){10}}{\thicklines
\put(100,250){\line(1,-1){10}}\put(110,240){\line(1,1){10}}
\put(120,250){\line(1,-1){10}}\put(130,240){\line(1,1){10}}
\put(140,250){\line(1,-1){10}} } \put(95,225){\vector(-3,-2){42}}
\put(155,225){\vector(3,-2){42}} \put(125,225){\vector(0,-1){30}}
\put(119,230){\scriptsize 6050}

\put(0,160){\line(1,0){50}}
\put(0,160){\line(0,1){10}}\put(50,160){\line(0,1){20}}
{\thicklines
\put(0,170){\line(1,-1){10}}\put(10,160){\line(1,1){10}}
\put(20,170){\line(1,-1){10}}\put(30,160){\line(1,1){20}} }
\multiput(40,170)(2,-2){5}{\circle*{.5}} \put(50,160){\circle*{3}}
\put(50,145){\vector(3,-2){32}} \put(25,145){\vector(0,-1){25}}
\put(19,150){\scriptsize 4226}

\put(100,160){\line(1,0){50}}
\put(100,160){\line(0,1){10}}\put(150,160){\line(0,1){20}}
{\thicklines
\put(100,170){\line(1,-1){10}}\put(110,160){\line(1,1){30}}
\put(140,190){\line(1,-1){10}} }
\multiput(120,170)(2,-2){5}{\circle*{.5}}
\multiput(130,160)(2,2){5}{\circle*{.5}}\multiput(140,170)(2,-2){5}{\circle*{.5}}
\put(130,160){\circle*{3}} \put(150,145){\vector(3,-2){32}}
\put(122,150){\scriptsize 625}

\put(200,160){\line(1,0){50}} \put(200,160){\line(0,1){10}}
\put(250,160){\line(0,1){20}} {\thicklines
\put(200,170){\line(1,1){20}}\put(220,190){\line(1,-1){10}}
\put(230,180){\line(1,1){10}}\put(240,190){\line(1,-1){10}}  }
\multiput(200,170)(2,-2){5}{\circle*{.5}}
\multiput(210,160)(2,2){5}{\circle*{.5}}
\multiput(220,170)(2,-2){5}{\circle*{.5}}
\multiput(230,160)(2,2){5}{\circle*{.5}}
\multiput(240,170)(2,-2){5}{\circle*{.5}}
\put(210,160){\circle*{3}} \put(217,150){\scriptsize 1430}

\put(0,80){\line(1,0){50}} \put(0,80){\line(0,1){10}}
\put(50,80){\line(0,1){40}}{\thicklines
\put(0,90){\line(1,-1){10}}\put(10,80){\line(1,1){40}} }
\multiput(20,90)(2,-2){5}{\circle*{.5}}
\multiput(30,80)(2,2){10}{\circle*{.5}} \put(30,80){\circle*{3}}
\put(25,65){\vector(0,-1){20}} \put(21,70){\scriptsize 159}

\put(100,80){\line(1,0){50}} \put(100,80){\line(0,1){10}}
\put(150,80){\line(0,1){40}}{\thicklines
\put(100,90){\line(1,1){20}}\put(120,110){\line(1,-1){10}}
\put(130,100){\line(1,1){20}} }
\multiput(100,90)(2,-2){5}{\circle*{.5}}
\multiput(110,80)(2,2){5}{\circle*{.5}}
\multiput(120,90)(2,-2){5}{\circle*{.5}}
\multiput(130,80)(2,2){10}{\circle*{.5}} \put(110,80){\circle*{3}}
\put(121,70){\scriptsize 704}

\put(200,80){\line(1,0){50}} \put(200,80){\line(0,1){10}}
\put(250,80){\line(0,1){40}}{\thicklines
\put(200,90){\line(1,1){40}}\put(240,130){\line(1,-1){10}} }
\multiput(200,90)(2,-2){5}{\circle*{.5}}
\multiput(210,80)(2,2){15}{\circle*{.5}}
\multiput(240,110)(2,-2){5}{\circle*{.5}}
\put(210,80){\circle*{3}} \put(223,70){\scriptsize 55}

\put(0,0){\line(1,0){50}} \put(0,0){\line(0,1){10}}
\put(50,0){\line(0,1){60}}{\thicklines \put(0,10){\line(1,1){50}}
} \multiput(0,10)(2,-2){5}{\circle*{.5}}
\multiput(10,0)(2,2){20}{\circle*{.5}} \put(10,0){\circle*{3}}
\put(21,-10){\scriptsize 11}
\end{picture}
\\[5mm]
\begin{picture}(80,220)(-80,0)
\put(-118,210){\underline{$1$-orbit, quartet, $\mbox{sum over
orbit} = m_5^{(c)} \times
p_6^{(b)}\left(\rule{0pt}{4mm}\right.$\phantom{\hspace{45mm}}}}
\put(120,208){\line(1,0){60}}\put(120,208){\line(0,1){20}}
\put(120,228){\line(1,-1){20}}\put(140,208){\line(1,1){10}}
\put(150,218){\line(1,-1){10}}\put(160,208){\line(1,1){10}}
\put(170,218){\line(1,-1){10}}
\put(182,210){$\left.\rule{0pt}{4mm} \right) = 11 \times 676$}

\put(0,160){\line(1,0){50}} \put(0,160){\line(0,1){30}}
\put(50,160){\line(0,1){20}} {\thicklines
\put(0,190){\line(1,-1){20}}\put(20,170){\line(1,1){10}}
\put(30,180){\line(1,-1){10}}\put(40,170){\line(1,1){10}}  }
\put(50,145){\vector(3,-2){32}} \put(25,145){\vector(0,-1){25}}
\put(19,150){\scriptsize 4226}

\put(0,80){\line(1,0){50}} \put(0,80){\line(0,1){30}}
\put(50,80){\line(0,1){40}}{\thicklines
\put(0,110){\line(1,-1){20}}\put(20,90){\line(1,1){30}} }
\multiput(30,100)(2,-2){5}{\circle*{.5}}\multiput(40,90)(2,2){5}{\circle*{.5}}
\put(40,80){\circle*{3}} \put(25,65){\vector(0,-1){25}}
\put(21,70){\scriptsize 765}

\put(100,100){\line(1,0){50}} \put(100,100){\line(0,1){10}}
\put(150,100){\line(0,1){20}}{\thicklines
\put(100,110){\line(1,-1){10}}\put(110,100){\line(1,1){20}}
\put(130,120){\line(1,-1){10}}\put(140,110){\line(1,1){10}}}
\multiput(100,80)(2,0){25}{\circle*{.5}}
\multiput(100,80)(0,2){11}{\circle*{.5}}
\multiput(150,80)(0,2){11}{\circle*{.5}}
\multiput(110,100)(2,-2){5}{\circle*{.5}}
\multiput(120,90)(2,2){5}{\circle*{.5}}
\multiput(130,100)(2,-2){5}{\circle*{.5}}
\multiput(140,90)(2,2){5}{\circle*{.5}} \put(120,80){\circle*{3}}
\put(119,70){\scriptsize 2286}

\put(0,10){\line(1,0){50}} \put(0,10){\line(0,1){10}}
\put(50,10){\line(0,1){40}}{\thicklines
\put(0,20){\line(1,-1){10}} \put(10,10){\line(1,1){40}}}
\multiput(0,-10)(2,0){25}{\circle*{.5}}
\multiput(0,-10)(0,2){11}{\circle*{.5}}
\multiput(50,-10)(0,2){11}{\circle*{.5}}
\multiput(10,10)(2,-2){5}{\circle*{.5}}
\multiput(20,0)(2,2){15}{\circle*{.5}} \put(20,-10){\circle*{3}}
\put(23,-20){\scriptsize 159}
\end{picture}
\\[5mm]
\end{array}
$

\lb{FigB} \caption{\bf Examples of maximal right orbits for size
$L=5_{_{_{}}}$ Anchored Cross paths. \hspace{23mm}\rm Notations
are the same as on Fig.6. Numbers standing below configurations
are the corresponding stationary state coefficients of the model C.
On the last picture two configurations suffer from the total
avalanche (\ref{w-add}). }
\end{figure}
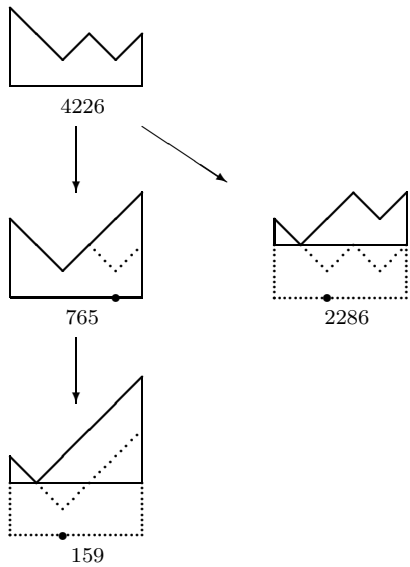

\begin{table}
\hfil
\begin{tabular}{ccccccccccccccccccccccccccccc}
&&
\begin{picture}(0,0)
\multiput(40,-3)(48,-22){3}{\line(2,-1){33}}
\put(26,5){\line(-2,1){25}}
\put(7,20){$\scriptstyle m=2n+3$}
\end{picture}
&&
$\lefteqn{\scriptstyle 0}$&&
$\lefteqn{\scriptstyle 0}$&&
$\lefteqn{\scriptstyle 0}$&&
$\lefteqn{\scriptstyle 0}$&&
$\lefteqn{\scriptstyle 0}$
\begin{picture}(0,0)
\multiput(16,19)(0,-47){2}{\line(0,-1){33}}
\put(18,20){$\scriptstyle n=2m+3$}
\end{picture}
&&
$\lefteqn{\!\!\scriptstyle { 429\over 32}}$&&
$\lefteqn{\!\!\scriptstyle  {143\over 32}}$&&
$\lefteqn{\!\!\!\!\!\scriptstyle  - {22\over 32}}$&&
\begin{picture}(0,0)
\put(18,19){\line(-2,-3){7}}
\multiput(1.5,-5.5)(-16.1,-23){13}{\line(-2,-3){7}}
\put(-208,-305){\line(-2,-3){10}}
\put(-245,-324.5){$\scriptstyle m=-1$}
\end{picture}
$\lefteqn{\scriptstyle   {1\over 8}}$&&
\begin{picture}(0,0)
\put(18,19){\line(-2,-3){7}}
\multiput(1.5,-5.5)(-16.1,-23){13}{\line(-2,-3){7}}
\put(-208,-305){\line(-2,-3){10}}
\put(-215,-324.5){$\scriptstyle m=0$}
\end{picture}
$\lefteqn{{\!\scriptstyle   {1\over 16}}}$&&
$\lefteqn{\!\scriptstyle   {14\over 32}}$&&&&
\\[3mm]
&&&
$\lefteqn{\!\!\!\!\!\scriptstyle { 252586\over 16}}$&&
$\lefteqn{\!\!\!\!\scriptstyle { 26194\over 16}}$&&
$\lefteqn{\scriptstyle 0}$&&
$\lefteqn{\scriptstyle 0}$&&
$\lefteqn{\scriptstyle 0}$&&
$\lefteqn{\scriptstyle 0}$&&
$\lefteqn{{\!\scriptstyle   {66\over 16}}}$&&
$\lefteqn{{\!\scriptstyle   {18\over 16}}}$&&
$\lefteqn{{\!\!\!\!\scriptstyle  - {1\over 8}}}$&&
$\lefteqn{{\!\!\!\!\scriptstyle  - {1\over 8}}}$&&
$\lefteqn{{\!\scriptstyle   {12\over 16}}}$&&
$\lefteqn{{\!\!\!\!\!\scriptstyle  - {84\over 16}}}$&&&
\\[3mm]
&&
$\lefteqn{\!\!\!\!\scriptstyle { 22610\over 8}}$&&
$\lefteqn{\!\!\!\scriptstyle { 8398\over 8}}$&&
$\lefteqn{\!\!\!\scriptstyle { 2431\over 8}}$&&
$\lefteqn{\!\!\scriptstyle { 429\over 8}}$&&
$\lefteqn{\scriptstyle 0}$&&
$\lefteqn{\scriptstyle 0}$&&
$\lefteqn{\!\scriptstyle { 21\over 8}}$&&
$\lefteqn{\scriptstyle { 7\over 8}}$&&
$\lefteqn{{\!\!\!\!\scriptstyle  - {1\over 4}}}$&&
$\lefteqn{{\scriptstyle   {1\over 4}}}$&&
$\lefteqn{{\!\!\!\!\scriptstyle  - {5\over 8}}}$&&
$\lefteqn{{\!\scriptstyle   {49\over 8}}}$&&
$\lefteqn{{\!\!\scriptstyle   {588\over 8}}}$&&
\\[3mm]
&$\lefteqn{\!\!\scriptstyle { 646\over 4}}$&&
$\lefteqn{\!\!\scriptstyle { 425\over 4}}$&&
$\lefteqn{\!\!\scriptstyle { 260\over 4}}$&&
$\lefteqn{\!\!\scriptstyle { 143\over 4}}$&&
$\lefteqn{\!\scriptstyle { 66\over 4}}$&&
$\lefteqn{\!\scriptstyle { 21\over 4}}$&&
$\lefteqn{\scriptstyle 0}$&&
$\lefteqn{{\!\!\!\!\scriptstyle  - {5\over 4}}}$&&
$\lefteqn{{\!\!\!\!\scriptstyle  - {1\over 2}}}$&&
$\lefteqn{{\scriptstyle  {1\over 4}}}$&&
$\lefteqn{{\!\!\!\!\scriptstyle  - {4\over 4}}}$&&
$\lefteqn{{\!\!\!\!\!\scriptstyle  - {25\over 4}}}$&&
$\lefteqn{{\!\!\!\!\!\scriptstyle  - {70\over 4}}}$&&
$\lefteqn{{\!\!\!\!\scriptstyle  - {147\over 4}}}$&
\\[3mm]
$\lefteqn{\!\scriptstyle { 17\over 2}}$&&
$\lefteqn{\!\scriptstyle { 15\over 2}}$&&
$\lefteqn{\!\scriptstyle { 13\over 2}}$&&
$\lefteqn{\!\scriptstyle { 11\over 2}}$&&
$\lefteqn{\scriptstyle { 9\over 2}}$&&
$\lefteqn{\scriptstyle { 7\over 2}}$&&
$\lefteqn{\scriptstyle { 5\over 2}}$&&
$\lefteqn{\scriptstyle { 3\over 2}}$&&
$\lefteqn{\scriptstyle { 1\over 2}}$&&
$\lefteqn{{\!\!\!\!\scriptstyle  - {1\over 2}}}$&&
$\lefteqn{{\!\!\!\!\scriptstyle  - {3\over 2}}}$&&
$\lefteqn{{\!\!\!\!\scriptstyle  - {5\over 2}}}$&&
$\lefteqn{{\!\!\!\!\scriptstyle  - {7\over 2}}}$&&
$\lefteqn{{\!\!\!\!\scriptstyle  - {9\over 2}}}$&&
$\lefteqn{{\!\!\!\!\!\scriptstyle  - {11\over 2}}}$
\\[3mm]
\begin{picture}(0,0)
\multiput(-5,2)(31.5,0){14}{\line(1,0){19}}
\put(437,2){\line(1,0){30}}
\put(470,2){$\scriptstyle n= - 1$}
\end{picture}
&\lefteqn{$\scriptsize 1$}&&\lefteqn{$\scriptsize 1$}&&
\lefteqn{$\scriptsize 1$}&& \lefteqn{$\scriptsize 1 $}&&
\lefteqn{$\scriptsize 1$}&&\lefteqn{$\scriptsize 1$}&&
\lefteqn{$\scriptsize 1$}&&\lefteqn{$\scriptsize 1$}&&
\lefteqn{$\scriptsize 1$}&&\lefteqn{$\scriptsize 1$}&&
\lefteqn{$\scriptsize 1$}&&\lefteqn{$\scriptsize 1$}&&
\lefteqn{$\scriptsize 1$}&&\lefteqn{$\scriptsize 1$}&
\\[3mm]
\begin{picture}(0,0)
\multiput(13.5,2)(31.5,0){15}{\line(1,0){19}}
\put(477,2){$\scriptstyle n=0$}
\end{picture}
\scriptsize 1\,&\,&\scriptsize 1\,&\,&\scriptsize 1\,&\,&\scriptsize 1\,&\,&
\scriptsize 1\,&\,&\scriptsize 1\,&\,&\scriptsize 1\,&\,&\scriptsize 1\,&\,&
\scriptsize 1\,&\,&\scriptsize 1\,&\,&\scriptsize 1\,&\,&\scriptsize 1\,&\,&
\scriptsize 1\,&\,&\scriptsize 1\,&\,&\scriptsize 1\,
\\[3mm]
&$\lefteqn{\!\!\!\!\scriptstyle - 8}$&&$\lefteqn{\!\!\!\!\scriptstyle  - 7}$&&
$\lefteqn{\!\!\!\!\scriptstyle - 6}$&&$\lefteqn{\!\!\!\!\scriptstyle  - 5}$&&
$\lefteqn{\!\!\!\!\scriptstyle - 4}$&&$\lefteqn{\!\!\!\!\scriptstyle  - 3}$&&
$\lefteqn{\!\!\!\!\scriptstyle - 2}$&&$\lefteqn{\!\!\!\!\scriptstyle  - 1}$&&
$\lefteqn{\scriptstyle 0
\begin{picture}(0,0)
\multiput(3,-2)(48,-22){3}{\line(2,-1){33}}
\put(146,-69){\line(2,-1){15}}
\put(164,-76){$\scriptstyle m=2n-1$}
\multiput(-2.5,-3)(0,-47){3}{\line(0,-1){33}}
\put(-2.5,-143){\line(0,-1){15}}
\put(0,-163){$\scriptstyle n=2m-1$}
\end{picture}
}$
&&$\lefteqn{\scriptstyle 1}$&&
$\lefteqn{\scriptstyle 2}$&&$\lefteqn{\scriptstyle 3}$&&
$\lefteqn{\scriptstyle 4}$&&$\lefteqn{\scriptstyle 5}$&
\\[3mm]
$\lefteqn{\!\!\!\!\!\!\scriptstyle -195}$&&$\lefteqn{\!\!\!\!\!\!\scriptstyle -132}$&&
$\lefteqn{\!\!\!\!\!\scriptstyle -84}$&&$\lefteqn{\!\!\!\!\!\scriptstyle -49}$&&
$\lefteqn{\!\!\!\!\!\scriptstyle -25}$&&$\lefteqn{\!\!\!\!\!\scriptstyle -10}$&&
$\lefteqn{\!\!\!\!\scriptstyle -2}$&&$\lefteqn{\scriptstyle 1}$&&
$\lefteqn{\scriptstyle 1}$&&$\lefteqn{\scriptstyle 0}$&&
$\lefteqn{\scriptstyle 0}$&&$\lefteqn{\scriptstyle 3}$&&
$\lefteqn{\scriptstyle 11}$&&$\lefteqn{\!\scriptstyle 26}$&&
$\lefteqn{\!\scriptstyle 50}$
\\[3mm]
&$\lefteqn{\!\!\!\scriptstyle 4356}$&&$\lefteqn{\!\!\!\scriptstyle 1764}$&&
$\lefteqn{\!\!\scriptstyle 588}$&& $\lefteqn{\!\!\scriptstyle 140}$&&
$\lefteqn{\!\scriptstyle 14}$&&$\lefteqn{\!\!\!\!\!\scriptstyle - 2}$&&
$\lefteqn{\scriptstyle 2}$&&$\lefteqn{\scriptstyle 2}$&&
$\lefteqn{\scriptstyle 0}$&&$\lefteqn{\scriptstyle 0}$&&
$\lefteqn{\scriptstyle 0}$&&$\lefteqn{\scriptstyle 0}$&&
$\lefteqn{\!\scriptstyle 26}$&&$\lefteqn{\!\!\scriptstyle 170}$&
\\[3mm]
&&$\lefteqn{\!\!\!\!\scriptstyle 39204}$&&
$\lefteqn{\!\!\!\scriptstyle 5544}$&&$\lefteqn{\!\!\scriptstyle 294}$&&
$\lefteqn{\!\!\!\!\!\scriptstyle - 18}$&&$\lefteqn{\scriptstyle 4}$&&
$\lefteqn{\!\!\!\!\scriptstyle - 2}$&&$\lefteqn{\scriptstyle 3}$&&
$\lefteqn{\scriptstyle 3}$&&$\lefteqn{\scriptstyle 0}$&&
$\lefteqn{\scriptstyle 0}$&&$\lefteqn{\scriptstyle 0}$&&
$\lefteqn{\scriptstyle 0}$&&$\lefteqn{\scriptstyle 0}$&&
\\[3mm]
&&&$\lefteqn{\!\!\!\!\!\!\!\!\scriptstyle -18018}$&&
$\lefteqn{\!\!\scriptstyle 528}$&& $\lefteqn{\!\!\!\!\!\scriptstyle -44}$&&
$\lefteqn{\scriptstyle 4}$&&$\lefteqn{\scriptstyle  2}$&&
$\lefteqn{\!\!\!\!\scriptstyle -8}$&&$\lefteqn{\!\!\!\!\!\scriptstyle -11}$&&
$\lefteqn{\scriptstyle 0}$&&$\lefteqn{\scriptstyle 0}$&&
$\lefteqn{\scriptstyle 0}$&&$\lefteqn{\scriptstyle 0}$&&
$\lefteqn{\scriptstyle 0}$&&&
\\[3mm]
&&&&$\lefteqn{\!\!\!\!\scriptstyle  -1716}$&&
$\lefteqn{\!\scriptstyle 52}$&&$\lefteqn{\scriptstyle 4}$&&
$\lefteqn{\scriptstyle 4}$&&$\lefteqn{\!\!\!\!\!\scriptstyle -10}$&&
$\lefteqn{\!\scriptstyle 26}$&&$\lefteqn{\!\scriptstyle 26}$&&
$\lefteqn{\scriptstyle 0}$&&$\lefteqn{\scriptstyle 0}$&&
$\lefteqn{\scriptstyle 0}$&&$\lefteqn{\scriptstyle 0}$&&
&&
\\[3mm]
&&&&&
$\lefteqn{\!\!\!\!\!\scriptstyle -60}$&& $\lefteqn{\!\!\!\!\scriptstyle -8}$&&
$\lefteqn{\!\!\!\!\scriptstyle -4}$&&$\lefteqn{\!\!\!\!\!\scriptstyle -12}$&&
$\lefteqn{\!\!\scriptstyle 100}$&&$\lefteqn{\!\!\scriptstyle 170}$&&
$\lefteqn{\scriptstyle 0}$&&$\lefteqn{\scriptstyle 0}$&&
$\lefteqn{\scriptstyle 0}$&&$\lefteqn{\scriptstyle 0}$&&
&&&
\\[12mm]
\end{tabular}
\hfil \caption{\small Particular solution (\ref{Saa}), (\ref{Sab})
of Pascal's hexagon relation (\ref{pascal2}). The numbers
standing along indicated lines are given by formulas
(\ref{simple1})--(\ref{simple4}). Knowing them one can reconstruct
data on the whole lattice by a repeated use of Pascal's
hexagon relation.} \lb{tab-S}
\end{table}

\begin{table}
\ba
\nonumber
\begin{array}{rclcrcl}
F_{4,2}&=&
2xy+y^2&
\hspace{7mm}&
F_{6,3}&=&
11 x^2y+12 xy^2+3y^3
\\[1mm]
F_{5,2}&=&
3x^2+6xy+2y^2&&
F_{7,3}&=&
26x^3+78x^2y+55xy^2+11y^3
\\[1mm]
F_{6,2}&=&
11x^2+12xy+3y^2&&
F_{8,3}&=&
170x^3+294x^2y+156xy^2+26y^3
\\[1mm]
F_{7,2}&=&
50x^2+30xy+5y^2&&
F_{9,3}&=&
646x^3+816x^2y+350xy^2+50y^3
\\[1mm]
F_{8,2}&=&
85x^2+42xy+6y^2&&
\\[1mm]
F_{9,2}&=&
133x^2+56xy+7y^2&&
\\[4mm]
F_{8,4}&=&
\lefteqn{170x^3y+294x^2y^2+156xy^3+26y^4}&&&
\\[1mm]
F_{9,4}&=&
\lefteqn{646x^4+2584x^3y+2839x^2y^2+1190xy^3+170y^4}&&&
\end{array}
\\[4mm]
\nonumber
\rule{145mm}{1pt}
\\[4mm]
\nonumber
\begin{array}{rclcrcl}
G_{2,2}&=& 2+5x+3x^2
&\hspace{7mm}&
G_{3,2}&=&3+7x+4x^2
\\[1mm]
G_{2,3}&=& 3+11x+11x^2
&\hspace{7mm}&
G_{3,3}&=&11+44x+59x^2+26x^3
\\[1mm]
G_{2,4}&=& 4+19x+26x^2
&\hspace{7mm}&
G_{3,4}&=&26+137x+255x^2+170x^3
\\[1mm]
G_{2,5}&=& 5+29x+50x^2
&\hspace{7mm}&
G_{3,5}&=& 50+321x+747x^2+646x^3\hspace{16.5mm}
\\[1mm]
G_{2,6}&=&6+41x+85x^2
&\hspace{7mm}&&&
\\[4mm]
G_{4,2}&=&
\lefteqn{-4x-9x^2-5x^3}&&&&
\\[1mm]
G_{4,3}&=&
\lefteqn{26+97x+121x^2+50x^3}&&&&
\\[1mm]
G_{4,4}&=&
\lefteqn{170+935x+1956x^2+1837x^3+646x^4}&&&&
\\[4mm]
G_{5,2}&=&
\lefteqn{5x+16x^2+17x^3+6x^4}&&&&
\\[1mm]
G_{5,3}&=&
\lefteqn{-50x-179x^2-214x^3-85x^4}&&&&
\\[4mm]
G_{6,2}&=&
\lefteqn{-6x-19x^2-20x^3-7x^4}&&&&
\end{array}
\ea
\caption{\small  Lattice polynomials $F_{m,n}(x,y)$ and $G_{m,n}(x)$
for small values of $m$ and $n$.}
\lb{tab-F}
\end{table}

\end{document}